\documentclass[times,authoryear,longtitle,twocolumn,nopreprintline,3p]{elsarticle}

\usepackage{framed,multirow}

\usepackage{url}
\usepackage{xcolor}

\usepackage{hyperref}

\definecolor{newcolor}{rgb}{.8,.349,.1}

\journal{Electrical Engineering and Systems Science > Image and Video Processing}

\usepackage{amsmath,amssymb,amsfonts}
\usepackage{algorithmic}
\usepackage{graphicx}
\usepackage{textcomp}

\usepackage{mathrsfs}
\usepackage{gensymb}
\usepackage{microtype}

\usepackage[inline]{enumitem}

\usepackage[space,multidot]{grffile}
\usepackage{subcaption}
\usepackage{svg}
\usepackage{booktabs}
\usepackage{siunitx}
\usepackage{url}
\usepackage{soul}

\usepackage{tikz}
\usetikzlibrary{shapes,arrows,positioning,calc}

\graphicspath{{./images/}}
\newcommand\myscale{0.1}

\begin{document}
% \verso{Vianna and Murta Jr.}

\begin{frontmatter}

\title{Long-range medical image registration through generalized mutual information (GMI): toward a fully automatic volumetric alignment}

\author{Vinicius Pavanelli Vianna\corref{cor1}}
\ead{vnpavanelli@wexperts.com.br}
\cortext[cor1]{Corresponding author}
\author{Luiz Otavio Murta Jr.\fnref{fn1}}
\ead{murta@usp.br}
\address{Department of Computing and Mathematics and Department of Physics, University of Sao Paulo, Ribeirao Preto, Brazil}

% \received{N/A}
% \finalform{N/A}
% \accepted{N/A}
% \availableonline{N/A}
% \communicated{V. Vianna}

% \maketitle

\begin{abstract}
Image registration is a key operation in medical image processing, allowing a plethora of applications. Mutual information (MI) is consolidated as a robust similarity metric often used for medical image registration.  Although MI provides a robust medical image registration, it usually fails when the needed image transform is too big due to MI local maxima traps. In this paper, we propose and evaluate a generalized parametric MI as an affine registration cost function. We assessed the generalized MI (GMI) functions for separable affine transforms and exhaustively evaluated the GMI mathematical image seeking the maximum registration range through a gradient descent simulation. 
We also employed Monte Carlo simulation essays for testing translation registering of randomized T1 versus T2 images.
GMI functions showed to have smooth isosurfaces driving the algorithm to the global maxima. 
Results show significantly prolonged registration ranges, avoiding the traps of local maxima. 
We evaluated a range of [-150mm,150mm] for translations, [-180\si{\degree},180\si{\degree}]  for rotations, [0.5,2] for scales, and [-1,1] for skew with a success rate of 99.99\%, 97.58\%, 99.99\%, and 99.99\% respectively for the transforms in the simulated gradient descent.
We also obtained 99.75\% success in Monte Carlo simulation from 2,000 randomized translations trials with 1,113 subjects T1 and T2 MRI images.
The findings point towards the reliability of GMI for long-range registration with enhanced speed performance.
\end{abstract}


\end{frontmatter}

\section{Introduction}
\label{sec:introduction}
Mutual information (MI) is a robust statistical function widely used to drive image registration. MI registration has many derived functions and algorithm techniques with the purpose of improving it. However, MI techniques keep the drawback of having local maxima, trapping, and slowing down the algorithm.  The local maxima problem is critical when one expects a fully automated registration process. Some medical applications, such as medical atlas creation from numerous images and image-guided therapies (IGT), demand long-range automated registration. \citep{Lancaster2000,Markelj2012,Pouliot2005} Furthermore, one could use an effective long-range registration to train a deep-learning-based registration \cite{Litjens2017}.

\subsection{History}
\label{sec:intro_history}
Created as common channel information by \cite{Shannon1963}, MI was first transposed to the image registration problem in \cite{Viola1997}. The first algorithm, named EMMA, was evaluated using 2D MRI slices. Registration was successful in 50 randomly generated transforms displaced by 32 pixels, rotations of 28\si{\degree}, and scalings up to 20\%.  \cite{Wells1996} obtained success rate from 100\% for 10mm translations and 10\si{\degree} rotations, dropping to 90\% for  20mm and 20\si{\degree}, 68\% for 20mm and 68\si{\degree} and finally 41\% for 100mm and 20\si{\degree} concerning MR-CT registration. 

\cite{Collignon1995a} showed similar results, i.e., up to 2.5 cm translations and 11\si{\degree} rotations recovered fully automatically. \cite{Maes1997} reported registrations for up to 10 \si{\degree} rotations and 40 mm translations by tunning the parameter optimization order.

\subsection{Derived Metrics}
\label{sec:intro_derived}
\cite{Studholme1999} introduced the normalized mutual information (NMI) in 1999 with success up to 30mm and 30\si{\degree} when using smaller fields of view. \cite{Rueckert2000} proposed higher-order mutual information to use spatial information but without details of the capture range the experiments used. \cite{Pluim2000} introduced combined mutual information and gradient information to add spatial information to MI, showing good results for the 15mm and 15\si{\degree} transform. Further improvements were achieved for MR-T1/PET registration, although those improvements did not reflect a median or maximum error in most applications.

\cite{Russakoff2004} introduced the regional mutual information (RMI) in 2004 and compared results to \cite{Pluim2000,Rueckert2000} and classical MI. The RMI showed to be more robust with a success rate above $85\%$ in target registration errors (TRE) with transforms up to 15mm. The study approached RMI as translation cost function showing good results in a range up to 60mm but with an abnormal unsymmetrical shape. \cite{Loeckx2010} introduced conditional mutual information (cMI), which improves the MI results for nonrigid B-spline registration. However, the algorithm demands up to 37 times more computational time than MI. \cite{Zhuang2011} introduced spatially encoded mutual information (SEMI) and tested it with free form deformation improving the robustness in non-uniformity fields compared to NMI.

\cite{Cheah2012} introduced a combination of intensity and gradient information (CMI) with excellent results for up to 60mm and 60\degree, and good results as CMI and MI performed well, but NMI failed some registrations. \cite{Pradhan2016} introduced enhanced mutual information (EMI) using ranges up to 10mm and 20\si{\degree} with better results than NMI and RMI. However, their measure as a cost function for translation showed a local maximum for the x-axis.

The MI function was also modified to use generalized entropy, introducing the entropic index $q$ that modifies the entropy behavior, becoming Shannon entropy as $q \to 1$.  \cite{Martin2004} used 2D images with transforms up to 10mm and $15\si{\degree}$ and $0.5 \leq q \leq 0.9$.  \cite{Sun2007} used 3D images with transforms up to 20mm and 5\si{\degree} and $0.3 \leq q \leq 0.9$. \cite{Mohanalin2010} used mammograms with transforms up to 5mm and 3\si{\degree} and $ q < 1 $. \cite{Khader2011} used 3D images with transforms up to 40mm and $40\si{\degree}$ and $ q = 2$. \cite{AmaralSilva2014} used 3D images with transforms up to 12mm and $15\si{\degree}$, and $0.1 \leq q \leq 3$.

Over the past two decades, existing published investigations assessing MI corroborate it as a valid and robust image registration technique. However, the local maxima and a restricted capture range still challenge MI as automatic registration contributing to human intervention free process withdrawing pre-alignment or post-processing to correct failed registrations.

\subsection{Paper proposal}
In this paper, we comparativelly analyzed the MI function using Shannon, Tsallis (GMI), and Mattes \citep{Mattes2001} MI entropy function implemented in the ITK software \citep{McCormick2014} in comparisom to the proposed generalized mutual information as cost funcion.  The evaluated ranges were $[-150\text{mm}, 150\text{mm}]$ translation, $[-180\si{\degree},180\si{\degree}]$ rotation, $[0.5,2.0]$ scale, and $[-1,1]$ skew. We investigated the generalided information $q$ pamentric index in the range $[0-3]$.

\subsection{Mutual Information Mathematics}
The original idea of MI was defined by \cite{Shannon1963} as 
$  R = H(x) + H(y) - H(x,y) $, where $R$ is the bit rate of a communications channel, $H(x)$ and $H(y)$ are the entropies of the two sources, $x$ and $y$, and $H(x,y)$ is the sources joint shared entropy. 
Shannon also defined the entropy as $H(x) = - \sum_i p(i) \log p(i)$ 
and the joint entropy as $H(x,y) = - \sum_{i,j} p(i,j) \log p(i,j)$, where $i$ and $j$ are events and $p(i)$ are the occurrence probability of the event $i$.

\subsection{Generalized Entropy}
\cite{Tsallis1988} proposed a generalization of Boltzmann-Gibbs statistics that is applicable to Shannon entropy and the MI concept by replacing the entropy to Tsallis entropy: 
$H_q \equiv k\frac{1-\sum_i p(i)^q}{q-1}$,
where $k$ is a conventional positive constant and $q$ is the entropic index.
Using the assumption of $k=-1$, from Shannon, we finally get 
$H_q(x) = \frac{1-\sum_i p(i)^q}{1-q}$, 
noting that $ \lim_{q \to 1} H_q(x) = H(x) = - \sum_i p(i) \log p(i)$.
Due to the pseudo-additivity of Tsallis entropy, for two independent systems, the joint entropy is defined as \citep{Tsallis1994}: 
\begin{equation}
\label{eq:tsallis_additive}
H_q(A,B) = H_q(A) + H_q(B) + (1-q) H_q(A) H_q(B)
\end{equation}

\subsection{Registration process}
\begin{figure}
    \centering
		\resizebox{0.45\textwidth}{!}{   
			\tikzstyle{block} = [draw, fill=blue!25, rectangle, 
    minimum height=3em,  font=\footnotesize]
%\tikzstyle{sum} = [draw, fill=blue!20, circle, node distance=1cm]
\tikzstyle{input} = [coordinate]
\tikzstyle{output} = [coordinate]
%\tikzstyle{pinstyle} = [pin edge={to-,thin,black}]

%\begin{figure}
%   \caption{\label{diagrama_bloco}Block Diagram}
%\begin{tikzpicture}[auto, node distance=1cm,>=latex']
\begin{tikzpicture}[
auto, 
ultra thick, 
>=latex',
%transform canvas={scale=0.8}
]
\node [block, minimum height=2cm] (metrica) {Metric};
\node [coordinate, above=5mm of metrica.west] (metrica_fixa) {};
\node [coordinate, below=5mm of metrica.west] (metrica_movel) {};
% \node [input, left=of metrica.135] (input_fixa) {Fixa};
\node [block, right=of metrica] (otimizador) {Optimizer};
\node [output, right=2cm of otimizador] (saida) {};
%\node [block, left=2cm of metrica_movel] (interpolador) {Interpolator};
\node [block, left=2cm of metrica_movel
, text width=6em, align=center
] (tg) {Transform};

\node [input, left=2cm of tg] (input_movel) {};
%\node [input, left=of metrica_fixa] (aux_fixa) {};
\node [input] (input_fixa) at (metrica_fixa -| input_movel) {Fixed};
\node [coordinate] (aux_fixa) at (input_fixa -| tg.west) {};
\node [coordinate] (aux_fixa_interpolador) at (input_fixa -| tg.east) {};
\node [coordinate] (aux_saida) at ( $ (otimizador)!0.5!(saida) $ ) {};
\node [coordinate, below=of tg] (feed_tg) {};
\node [coordinate] (feed_saida) at (feed_tg -| aux_saida) {};
%\node at (input_fixa) [above = 1mm of input_fixa] {Fixa};
\draw[-] (input_fixa) -- node {Fixed} (aux_fixa);
\draw[-] (aux_fixa) -- (aux_fixa_interpolador);
\draw[->] (aux_fixa_interpolador) -- node {$u(x)$} (metrica_fixa);

%\draw[->] (tg) -- (interpolador);
\draw[->] (tg) -- node {$T(v(x))$} (metrica_movel);
\draw[->] (input_movel) -- node {Moving} (tg);
\draw[->] (metrica) -- node {$F_m(\cdot)$}(otimizador);
\draw[->] (otimizador) -- node {$\hat{T}$} (saida);
\draw[-] (aux_saida) -- (feed_saida);
\draw[-] (feed_saida) -- node {$T$} (feed_tg);
\draw[->] (feed_tg) -- (tg.south);

%\draw[green,thick,dotted,fill=green!25,fill opacity=0.2] (metrica.north |- aux_fixa_interpolador) rectangle (feed_saida);

% $ (otimizador)!0.5!(saida) $ 

%\node [coordinate] (vermelho1) at (aux_fixa_interpolador |- metrica.north) {};
\node [coordinate] (vermelho_aux) at (input_fixa |- metrica.north) {};
\node [coordinate] (vermelho1) at ($(vermelho_aux)!0.0!(metrica.north)$) {};
\node [coordinate] (vermelho2_aux) at (otimizador.west |- feed_saida) {};
\node [coordinate] (vermelho2) at ($(otimizador.west)!0.7!(vermelho2_aux)$) {};

\draw[red, ultra thick,dotted,fill=red!25,fill opacity=0.2] (vermelho1) rectangle (vermelho2);

\end{tikzpicture}
%\end{figure}
		}
  \caption{Basic image registration block diagram, \texttt{fixed} and \texttt{moving} are the images to be registered, represented by the functions $u()$ and $v()$, $T$ are the transformation parameters used and $\hat{T}$ are the final solution provided by the optimizer at the end of the process, $F_m(\cdot)$ is the metric output to the images.}
    \label{fig:block_diagram}
	\end{figure}
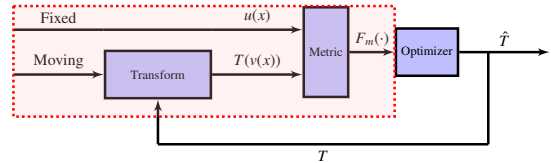

The first step in image registration with MI is understanding how it works. The basic registration process is represented in the block diagram of Fig. \ref{fig:block_diagram}.  We have two images, i.e., fixed and moving, where the fixed image is submitted to the similarity metric as is, and the moving image is transformed, using a set of parameters $T$, e.g., a set of translation vectors, and then submitted to the similarity metric. 

The similarity metric block analyses both image input and provides a scalar output that measures how the images are similar. This metric signal enters the optimizer block that processes it and generates a new set of parameters $T$ to transform, creating a new iteration step. The optimizer evaluates how the metric signal changes depending on the parameters $T$ on each iteration. In an ideal scenario, the optimizer produce a sequence of $T_i$ that progress towards the best metric signal, i.e., supposedly, the best image registration, this final set of parameters $T$ at registration end is called $\hat{T}$.

The image registration optimizer works driven by the similarity metric signal. It worth noting that an inefficient metric will produce inefficient signals, then the registration fails even with a good optimizer. 
The way to achieve a better registration, as we believe, is to check the metric signal output in relation to the input $T$, and not just trying a new metric and only checking the $\hat{T}$ output from it. A good metric signal drives the optimizer to a good registration.

One can see this approach in Fig. \ref{fig:block_diagram} as we first analyzed the red shaded region of the diagram by providing the images and checking how the metric changes as we change the transform parameters. We focus our interest on the $F_m(\cdot)$ output and not the $\hat{T}$ output, relying on the premise that a better $F_m(\cdot)$ space will lead to a better final $\hat{T}$ registration parameter.

\section{Materials and methods}
\subsection{Images}
We selected the images from two image databases, i.e., the Human Connectome Project (HCP) young adults \citep{Glasser2013,VanEssen2011}, and the NAMIC Registration Library.  One can find the detailed description and the acquisition protocols at the HCP website
\footnote{\url{http://protocols.humanconnectome.org/HCP/3T/imaging-protocols.html}}.

We selected the MRI 3T pre-processed structural images only from the HCP from patients number 172635 and 211821, with a size of $[256, 320, 320]$ and isotropic spacing of $0.7$mm. We also used images (N2\_T1, N2\_T2, and N4\_T1) from the NAMIC Registration Library Case 19\footnote{\url{https://www.na-mic.org/wiki/Projects:RegistrationLibrary:RegLib_C19}},
provided by the UNC Midas Database of healthy volunteers, with a size of $[176, 256, 176]$ and isotropic spacing of $1$mm. 
In special, Mattes MI initial tests were reduced from 3 days of computing time using HCP images to 1 day using UNC Midas.

\subsection{GMI additivity}
One can define two equations for GMI to compute the MI, we analyzed the both equations in this paper, i.e.:
\begin{align}
\label{eq:MI_normal}  I_q(x,y) &= H_q(x) + H_q(y) - H_q(x,y) \\
\label{eq:MI_additive} I_q(x,y) &= H_q(x) + H_q(y) - H_q(x,y) + (1-q)H_q(x)H_q(y) 
\end{align}

Equation \eqref{eq:MI_normal} is the original formulation made by Shannon.  In this paper, we will refer to \eqref{eq:MI_normal} as ``nonadditive'' GMI, while \eqref{eq:MI_additive} is the ``additive'' GMI. 
The additive GMI equation \eqref{eq:MI_additive} is used in previous Tsallis registration studies mentioned in section \ref{sec:introduction}.
However, equation  \eqref{eq:tsallis_additive} is the one defined for independent systems, where $I_q(x,y) = 0$, so using equation \eqref{eq:MI_additive} we can hold this property true for those systems. In order to succeed, MI registration needs mutual information between the images, i.e., $I_q(x,y) > 0$, showing an information dependency between the images as a requisite for MI registration. To the best of our knowledge, Tsallis nonadditive GMI, i.e., equation \eqref{eq:MI_normal}, was not used in the literature for image registration.

Furthermore, a correct GMI function for Tsallis entropy is a controversial question in the literature. As noted by \cite{Furuichi2006}, one should not use the term GMI with Tsallis entropy as ``proper evidence of channel coding theorem for information has not ever been shown in the context of Tsallis statistics''. \cite{Yamano2001} develops the GMI equation as: $I_q(Y;X) = [ H_q(X) + H_q(Y) - H_q(X,Y) + (q-1)H_q(X)H_q(Y) ] \times [1 + (q-1)H_q(X)]^{-1}$, which have an extra component in the denominator regarding equation \eqref{eq:MI_additive}. Finally, \cite{Sparavigna2015} shows that the Yamano equation is not symmetrical, i.e., $I_q(Y;X) \neq I_q(X;Y)$, and proposed a new equation:  $I_q(Y;X) = [ H_q(X) + H_q(Y) - H_q(X,Y) + (q-1)H_q(X)H_q(Y) ] \times [1 + (1-q) \text{max}\{H_q(X),H_q(Y)\}]^{-1}$, using the maximum between the two entropies at denominator. 

\subsection{Histogram binning}
Histogram binning is implemented by disabling the least significant bits of the images, in a similar way done by \cite{Collignon1995}. We converted all images voxels to 16 bits unsigned integers (\texttt{uint16\_t}), and then a simple bitwise \texttt{AND} operator applied the binning bitmask. The notation used in this paper is for the number of bits left in the image, so when we use 6 bits binning histogram, we left 6 bits on both images, using a bitmask of $1111.1100.0000.0000$, and the joint histogram will have $6\times6=12$ bits, resulting in 2048 grey levels. We first applied a normalization to spread the histogram to all 16 bits levels to improve the binning results: $ x' = 2^{16} \times (x-x_{min}) / (x_{max}-x_{min}) $, 
where $x'$ is the normalized grey value, $x$ is the input value and $x_{max}$, $x_{min}$ are the maximum and minimum values in the entire image.

\section{Experimental}
\subsection{Metric image visualization}
\label{sec:experimental_visualization}
To understand how the metric reacts to geometric transforms we mapped a set of transforms into a 3D output space, representing the parameters ranges: \begin{description*}[mode=unboxed, itemjoin={{; }}, afterlabel={{\nobreakspace}}, after={{.}}]
	%\begin{description*}[noitemsep,nosep]
	\item[translation] are mapped directly to a cube showing the $[-150mm,150mm]$ range in all axes 
	\item[rotation] are mapped to a cube with $[-1,1]$ range in all axes, with the transformation $x = \sin (\theta / 2)$ where $\theta$ is the angle and $x$ is the transformed parameter
	\item[scale] are mapped to a cube with $[-1,1]$ range in the three axes, with the transformation: $x = (1+s)^{-1} \,\text{if}\, s < 0$; $x = 1 \,\text{if}\, s = 0$; $x = (1+s) \,\text{if}\, s > 0$, where $s$ is the real scale and $x$ is the transformed parameter
	\item[skew] are mapped directly to a cube with $[-1,1]$ range in the three axes
\end{description*}

To create the graphs in the defined cube, i.e., transform range, we divided the 3D space in $51$ points, creating a cube with dimensions of $51 \times 51 \times 51$, where each voxel of this cube represent a single unique transformation parameter set, e.g. a translation of $[6\text{mm}, 12\text{mm}, 24\text{mm}]$ in the axes $x$, $y$ and $z$ respectively, would be mapped to point $[26,27,29]$, and the point $[25,25,25]$ would be the center, i.e., $[0\text{mm}, 0\text{mm}, 0\text{mm}]$ translation.

We believe the three-dimensional (3D) representation makes the visualization intuitive since the $x$ axis of the cube is related to transforms in the $x$ axis of the moving image. Finally, we set each voxel in the cube to the MI and GMI function value using the 
transformation parameters related to the voxel coordinates.

\subsection{3D visualization using isosurfaces}
\begin{figure*}
	\centering
	\subcaptionbox{}{\includegraphics[width=0.3\textwidth]{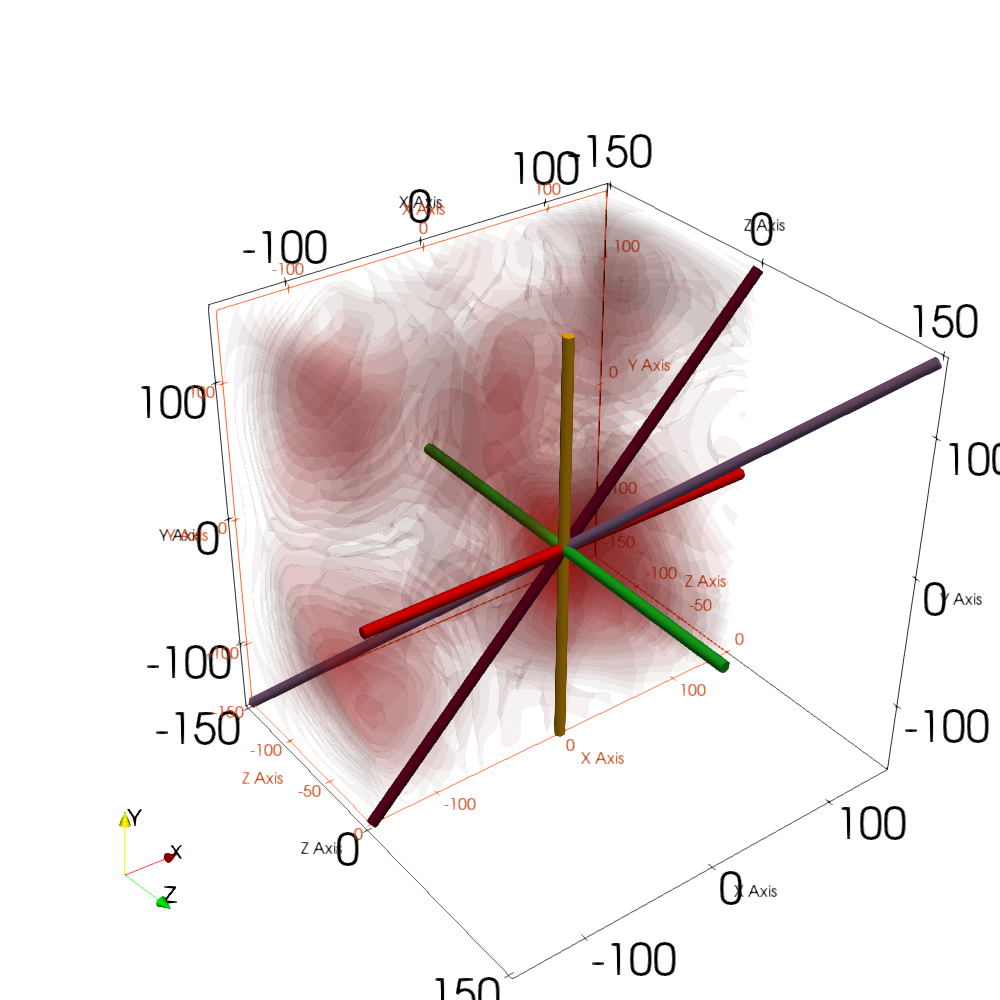}}%
	\hfill
	\subcaptionbox{}{\includegraphics[width=0.3\textwidth]{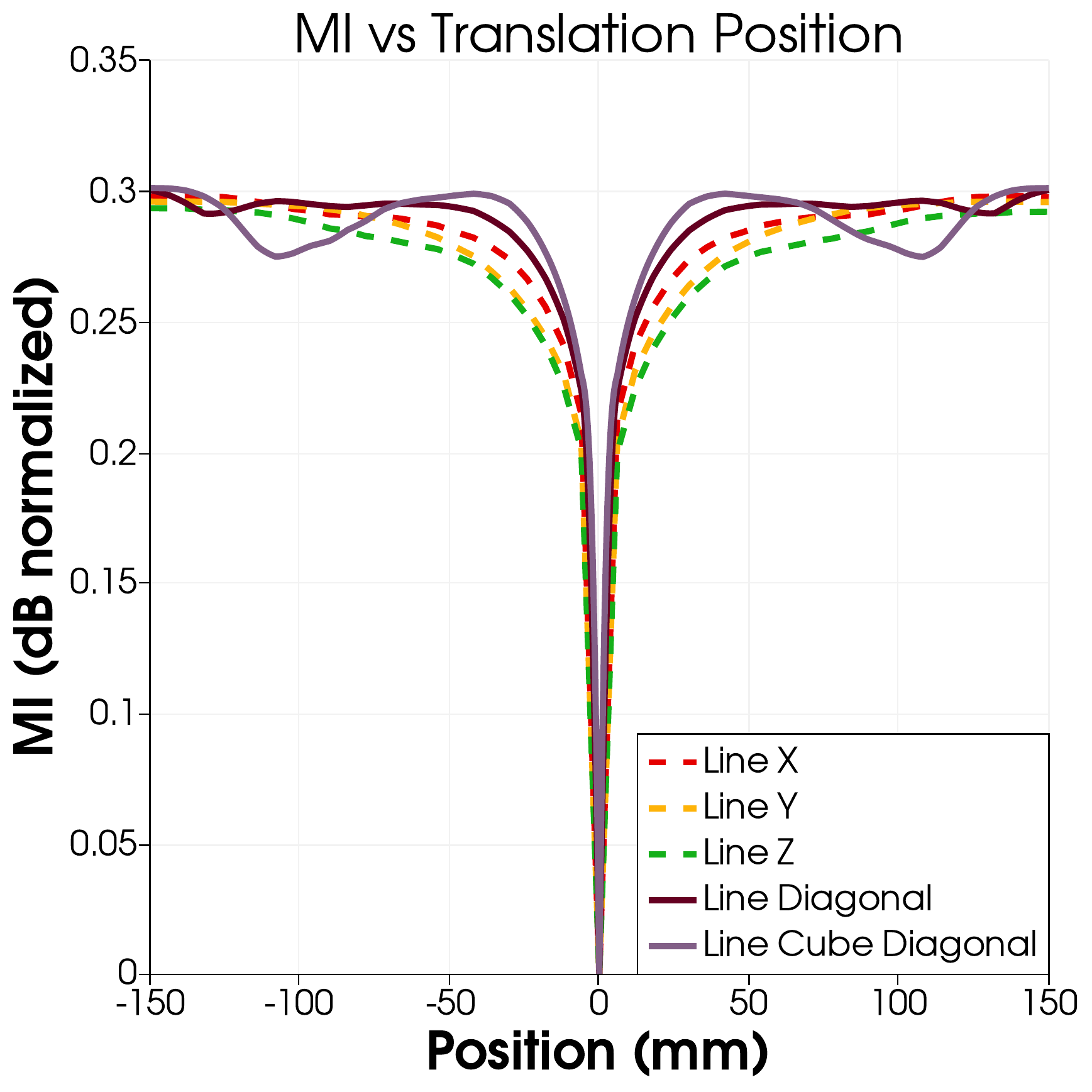}}%
	\hfill
	\subcaptionbox{}{\includegraphics[width=0.3\textwidth]{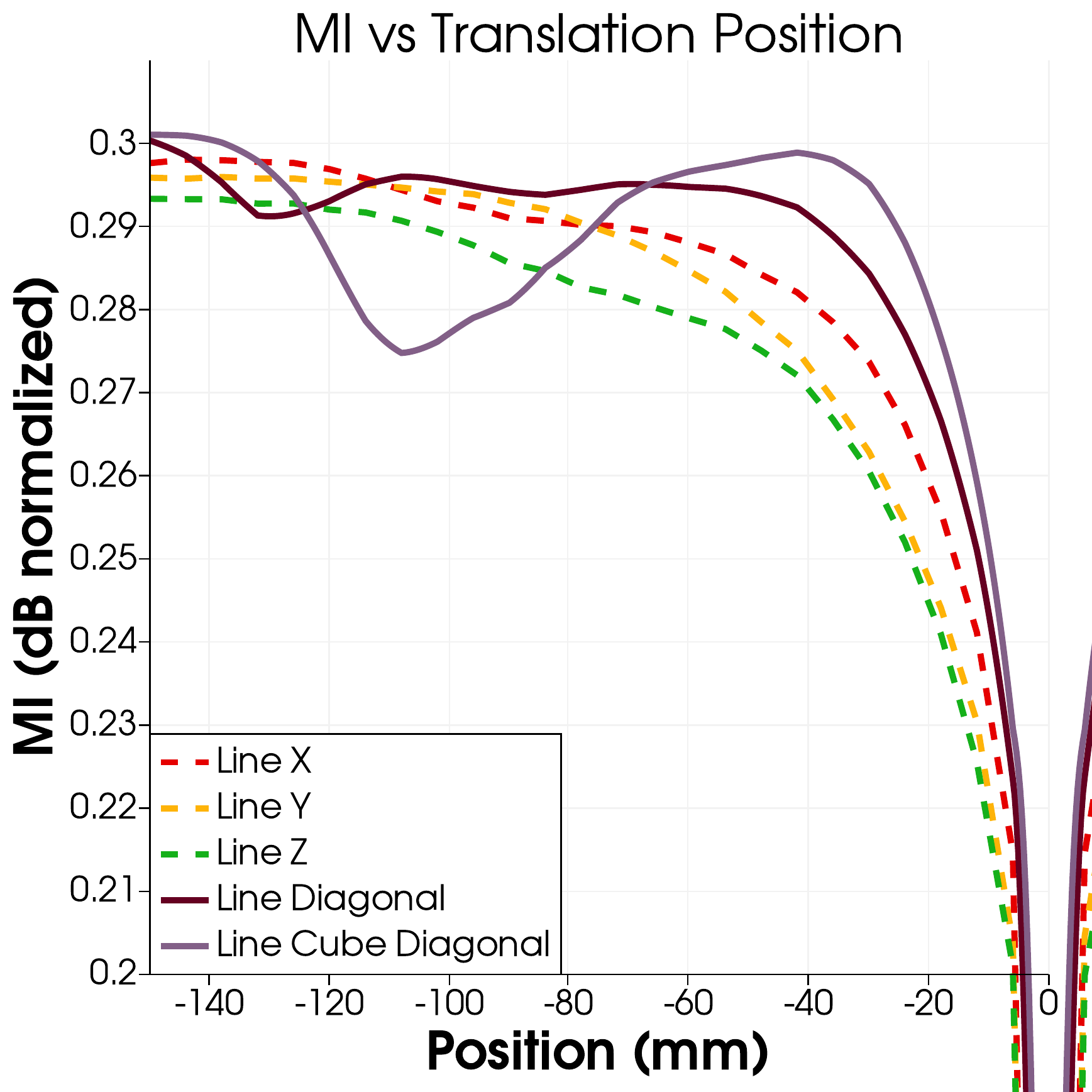}}%
	\hfill
	\caption{\label{mattes_plotlines}Relation between isosurfaces and lineplots: (a) isosurfaces of Mattes MI with lineplots probes in colors (colored lines in comparison to Fig. \ref{translation_surfaces}); (b) full 1D plots of probes; (c) detail of 1D plots to show local minima}
\end{figure*}

Fig. \ref{mattes_plotlines} introduces the 3D isosurfaces visualization used to represent the transform parameter ranges at Fig. \ref{mattes_plotlines}a. We show the MI profiles in 2D plots, as usual in the literature, in Fig. \ref{mattes_plotlines}b, and a zoom of the 2D plots in Fig. \ref{mattes_plotlines}c.  
Each tube in the isosurface 3D plot (Fig. \ref{mattes_plotlines}a) represents a pathway to plot the MI function profile (Fig. \ref{mattes_plotlines} b-c). We used tubes centered on the image, and one tube in a diagonal over plane $z=0$ and the other diagonal over the cube, from the minimum of all axes to their maximum. The dashed lines on the 2D plots are the usual axes in literature, and one can not see any local minima except the solution. The solid lines represent profiles taken along the diagonals, and one can see local minima. In the 3D image, we used an opacity effect to highlight the local minima present on the cube diagonal. The local minima are the structures with less opacity on the four corners.

Furthermore, the gradient lines are normal, i.e., perpendicular, to the isosurfaces. The most common MI optimization method is gradient descent, making this 3D plot analysis an intuitive way to understand how and why a gradient descent optimizer succeeds or fails.

\subsection{Registration Simulation}
\label{sec:registration_simulation}
Using the metric images from section \ref{sec:experimental_visualization}, we created a very naive algorithm that simulates a gradient descent registration. Since we are only interested, at this point, in how much of the space we can register with each metric, we use the center of the metric image as a seed, i.e., the gold standard in the simulation. We grow the registration space around this seed if the neighbors are greater than our seed since that would give a gradient pointing to the seed. The algorithm used does not guarantee the same results as in the real scenario since the gradient descent's adaptable learning rate is not simulated, i.e., it is equivalent to using a constant learning rate equal to the image spacing that is not used in real practice.

Despite the simulation's limitations, we can estimate, quantitatively, each metric's capacity of registration in the transformation scenario studied, e.g., a scenario with only the translation transform. Furthermore, this provides more quantitative value from the metric images generated from section \ref{sec:experimental_visualization}, and are very fast to calculate since the algorithm is straightforward.

\subsection{Monte Carlo}

We carried out Monte Carlo simulations by probing translation parameters using a normal distribution with a standard deviation of $50 mm$, giving a three-sigma region ($99.7\%$ of the transforms) within [$-150$mm, 150mm] in each translation axes. We carried out four scenarios for the image registrations, enumerated in Table \ref{table:monte_carlo_scenarios}, with the first two scenarios using a single subject image and the latter two scenarios using a set of randomized subjects from the HCP dataset in each registration.

These simulations provide additional reliable assessments than the ones from section \ref{sec:registration_simulation}, and the randomized scenarios simulate inter-subject and inter-modality clinical registrations. An essay consists of 1,000 (or more) registration trials, logging each registration start parameters, end parameters, fixed and moving images used, the time spent to register, and a flag if the registration is within 5mm of the center. Since the HCP images are rigidly registered, we used the center as a gold standard. We assumed that the already provided HCP registration error lies in this $5 mm$ range for inter-subject essays.

\begin{table}
  \centering
\begin{tabular}{lccl}
  \toprule
  Name & Fixed & Moving & Subject \\
  \midrule
  T1 & T1 & T1 & Same\\
  T2 & T1 & T2 & Same\\
  Randomized T1 & T1 & T1 & Randomized subject \\
  Randomized T2 & T1 & T2 & Randomized subject \\
  \bottomrule
\end{tabular}
\caption{Registration scenarios using Monte Carlo simulation}
\label{table:monte_carlo_scenarios}
\end{table}

\section{Results}

\subsection{Translation}

\begin{figure*}
	\centering
	%    \hfill
	\subcaptionbox{}{\includegraphics[width=0.24\textwidth]{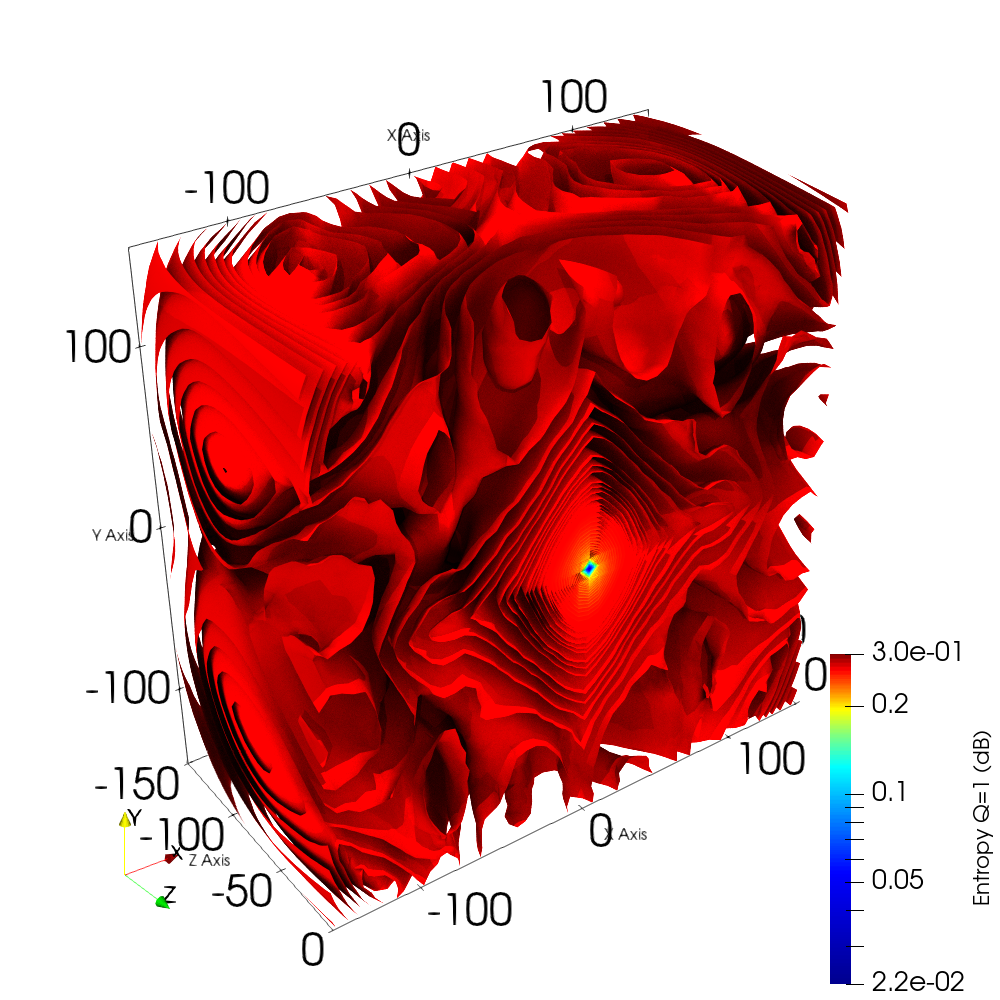}}%
	\hfill
	\subcaptionbox{}{\includegraphics[width=0.24\textwidth]{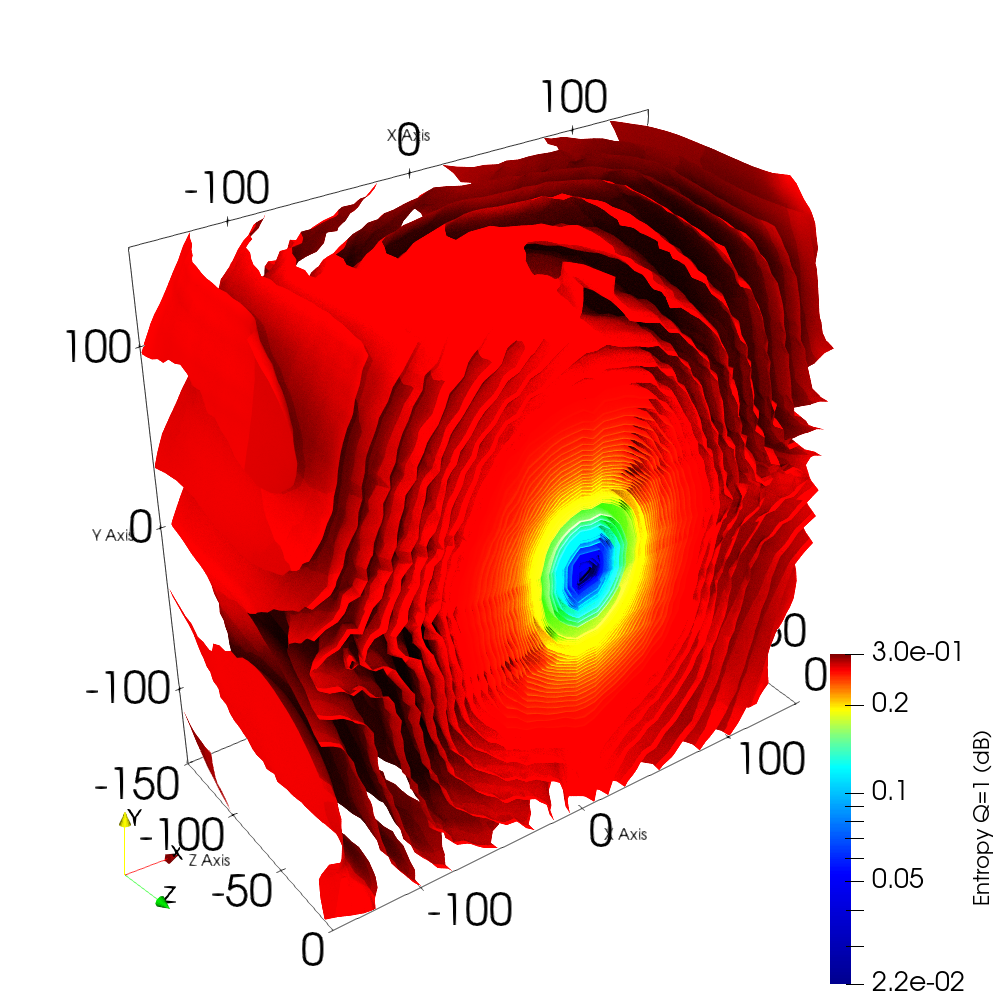}}%
	\hfill
	%    \hfill
	\subcaptionbox{}{\includegraphics[width=0.24\textwidth]{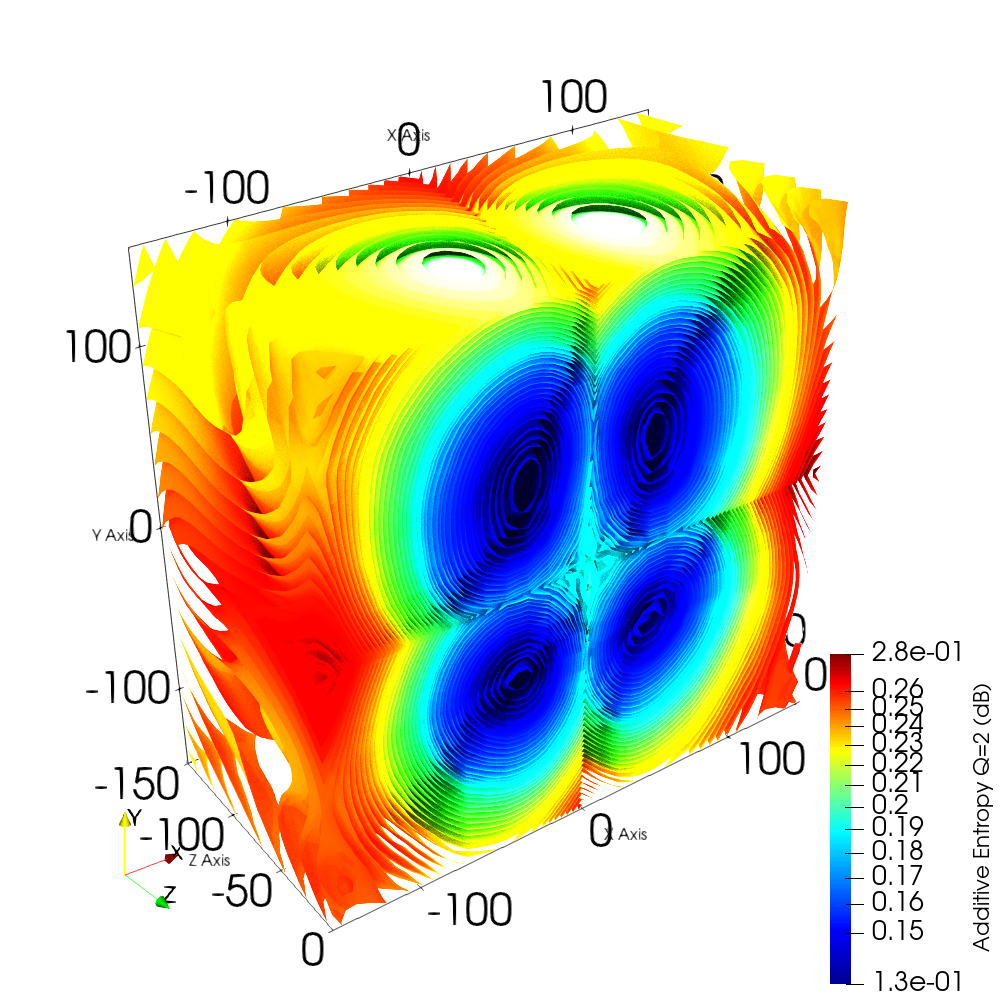}}%
	\hfill
	\subcaptionbox{}{\includegraphics[width=0.24\textwidth]{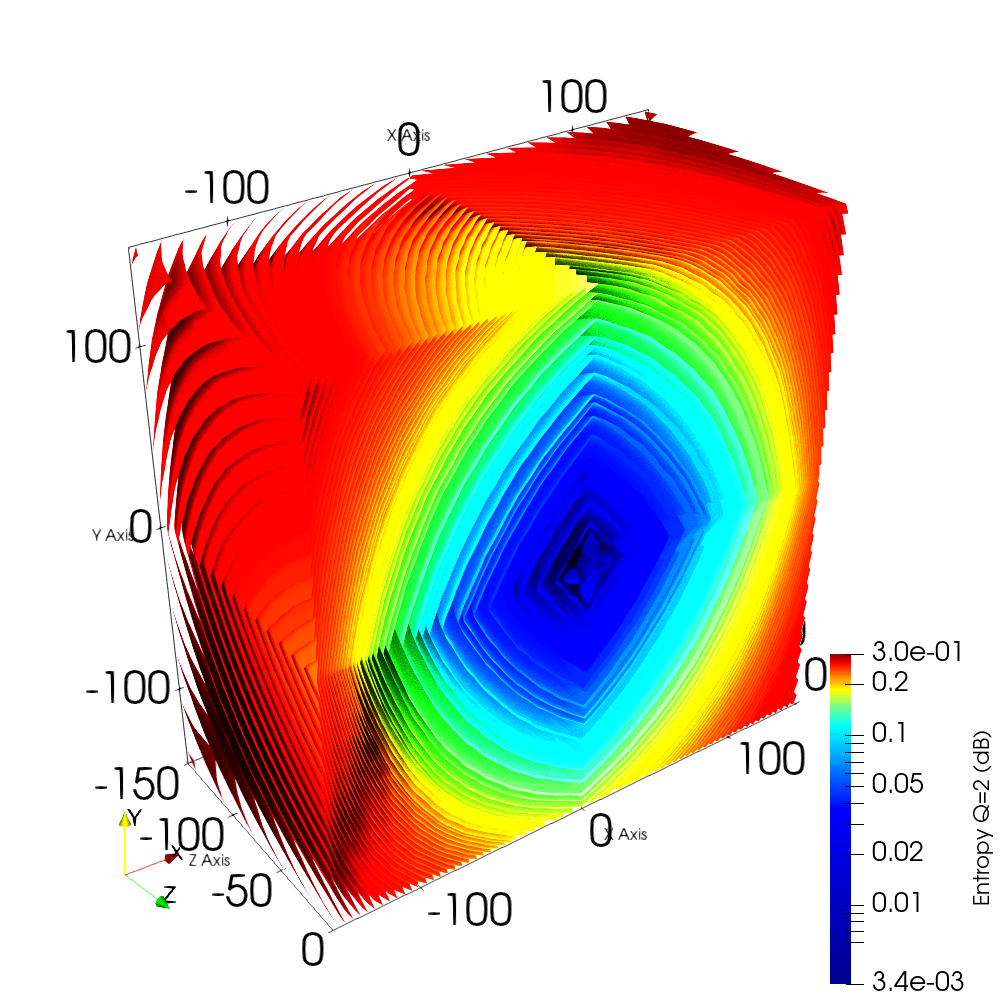}}%
	\hfill
	\caption{\label{translation_surfaces}Isosurfaces of the MI function using translation parameters: (a) Mattes, (b) Shannon, (c) Tsallis GMI Additive q=2.0, and (d) Tsallis GMI nonadditve q=2.0 }
\end{figure*}

Fig. \ref{translation_surfaces} shows the isosurfaces contour of the MI function when we use only the translation transformation on the HPC images. Mattes (Fig. \ref{translation_surfaces}a) have several local minima points that prevent long-range for translations registration, i.e., the existing local minima trap the optimization algorithm. Shannon MI does not have local maxima except for the solution, although it is challenging for the optimizer to work on nonsmooth surfaces. GMI has several local maxima when using the additive version. However, the nonadditive function has just the maximum solution, and very smooth surfaces provide excellent registration support.

The 3D isosurface plot explains the critical problem for MI as a cost function. The narrow registration range of some literature experiments due to the local minima inside the range. Besides, some Tsallis GMI experiments have instability since the previous studies in the literature use the additive Tsallis MI, plotted in Fig. \ref{translation_surfaces}c. At the same time, we have better performance using nonadditive Tsallis GMI, as seen in Fig. \ref{translation_surfaces}d.

\subsection{Rotation}
\begin{figure*}
	\centering
  \subcaptionbox{}{\includegraphics[scale=\myscale]{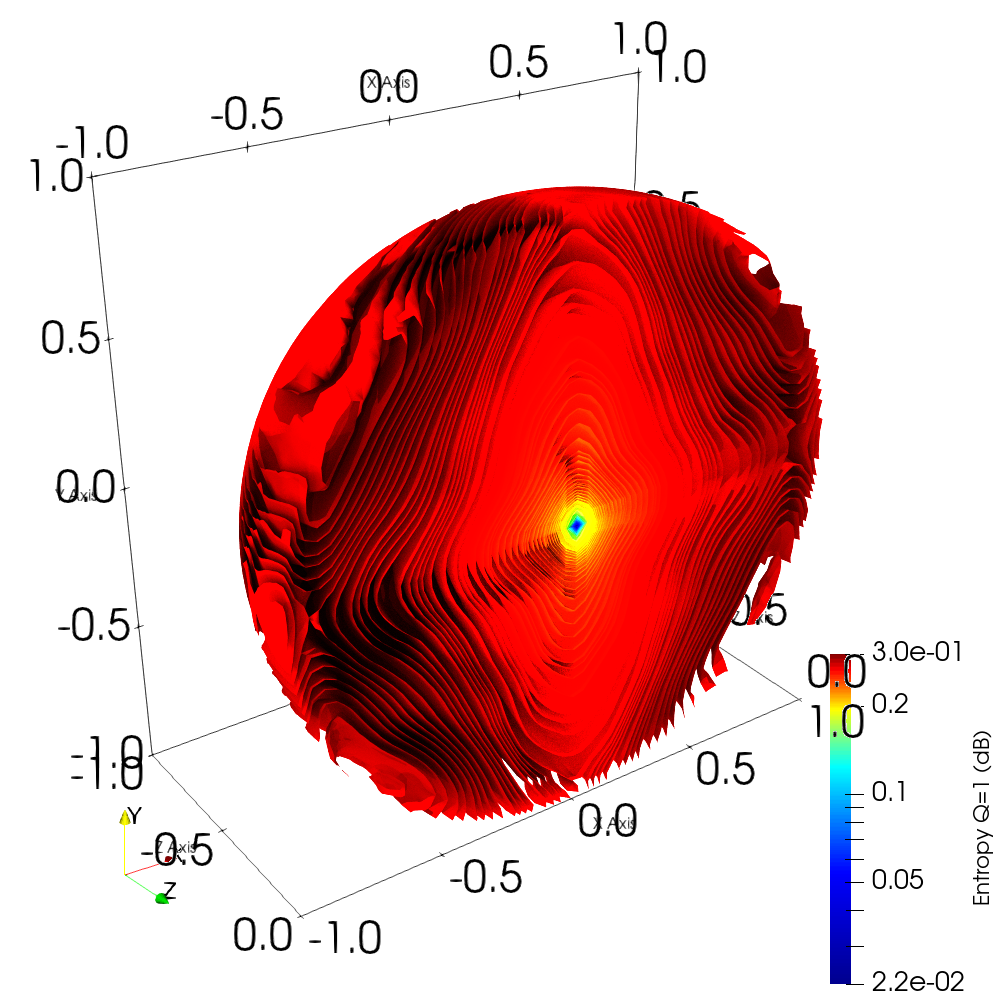}}
	\hfill
	\subcaptionbox{}{\includegraphics[scale=\myscale]{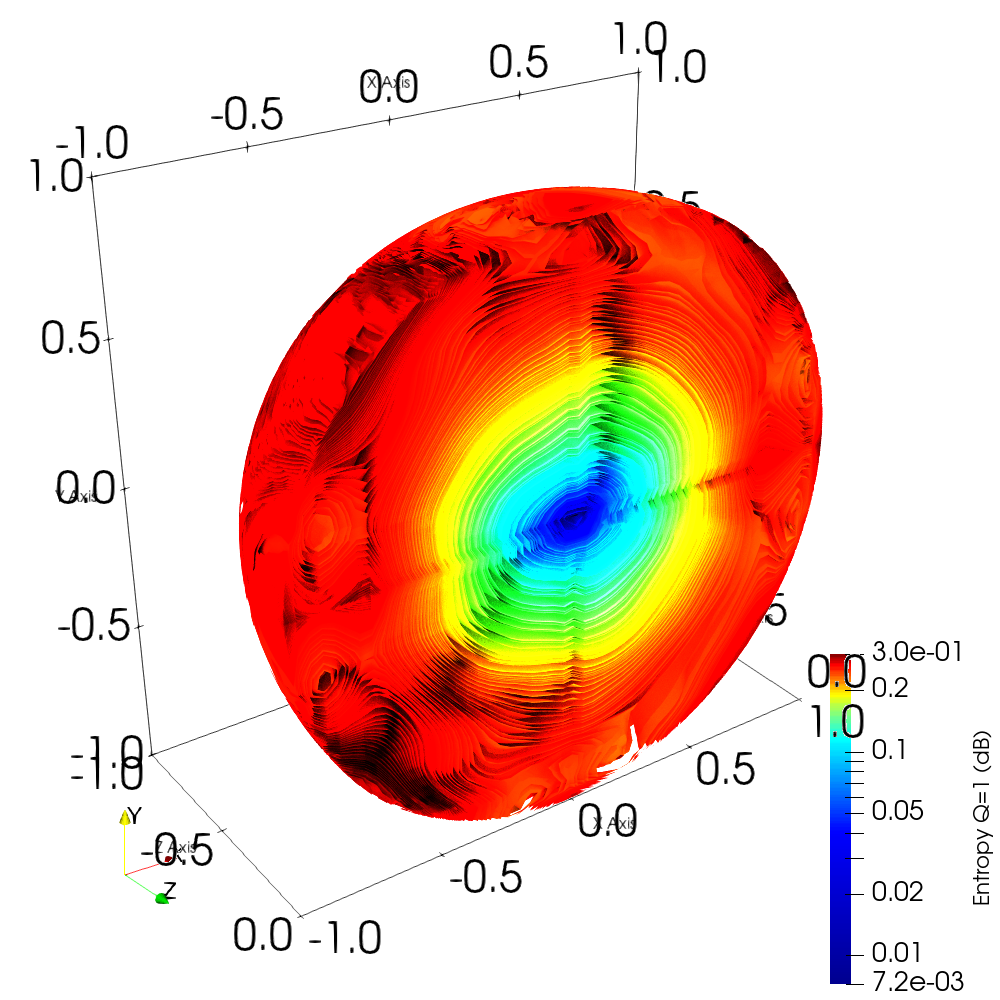}}%
	\hfill
	\subcaptionbox{}{\includegraphics[scale=\myscale]{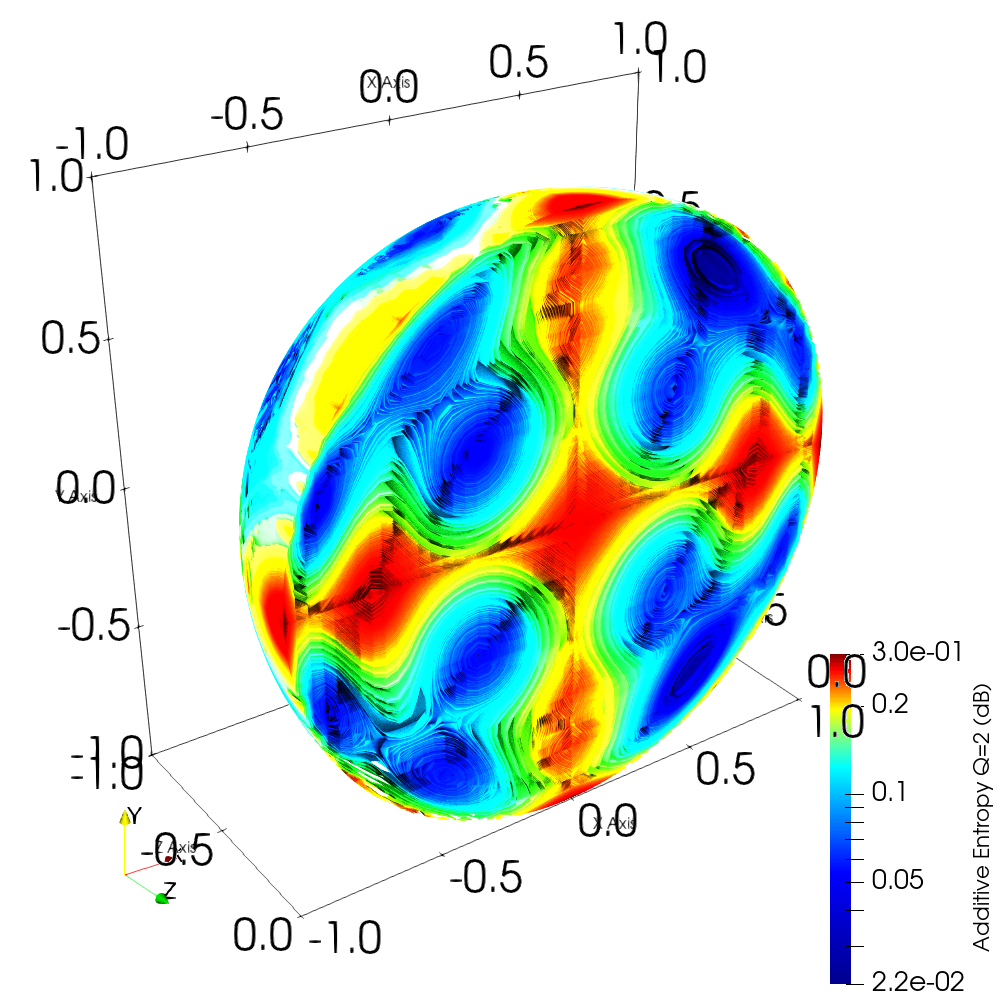}}%
	\hfill
	\subcaptionbox{}{\includegraphics[scale=\myscale]{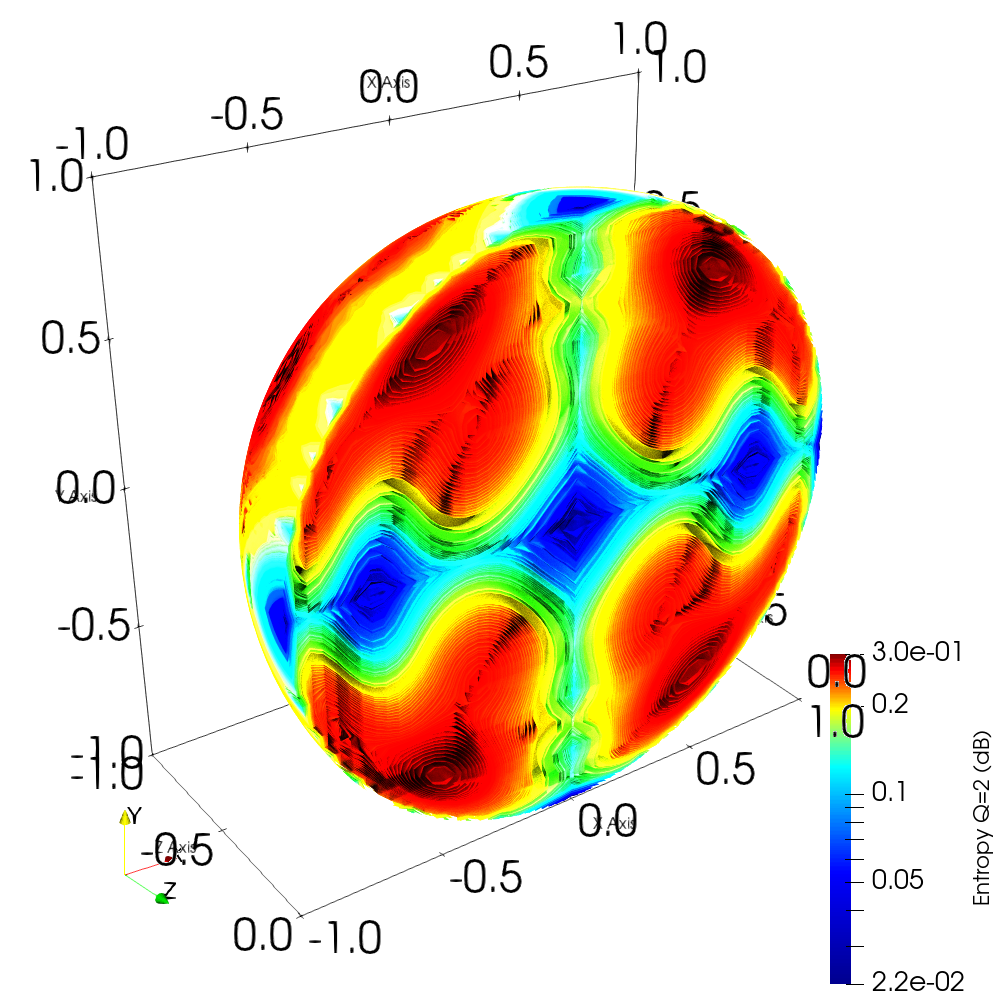}}%
	\hfill
	\subcaptionbox{}{\includegraphics[scale=\myscale]{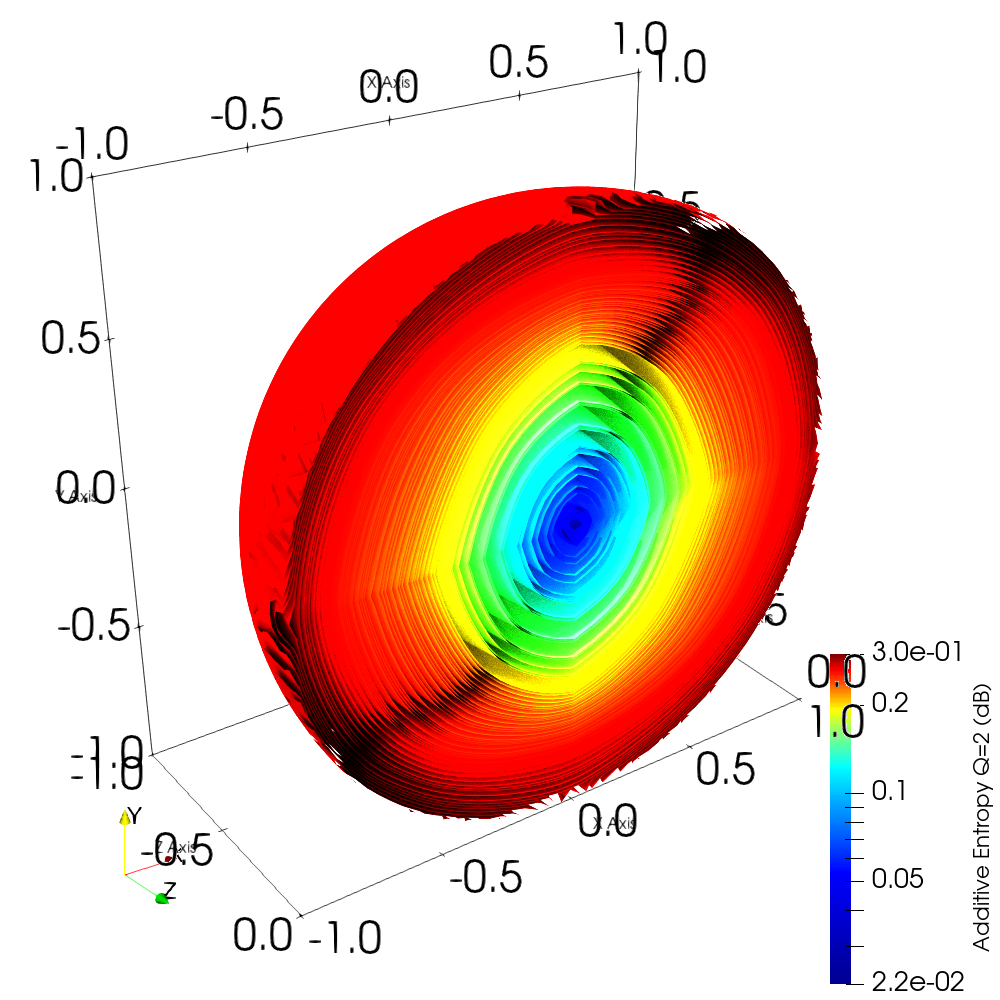}}%
	\caption{\label{rotation_surfaces}Isosurfaces of the MI function using rotation parameters: (a) Mattes, (b) Shannon, (c), Tsallis GMI additive q=2.0, (d) Tsallis GMI nonadditve q=2.0, and (e) Tsallis GMI additive q=2.0 with 6 bits bins}
\end{figure*}

Rotation transformation isosurfaces shown in Fig. \ref{rotation_surfaces} present different results. Mattes have minor local minima very far from the center, showing a wide registration range for rotation. In contrast, Shannon entropy also has small local maxima but closer to the center than Mattes. Tsallis GMI function has massive local extrema points and has some relationship between the additive and nonadditive function, with one being almost the other's negative, both imposing challenges to image registration. 

The main advantage of Mattes over Shannon and Tsallis GMI can be credited mostly to the histogram binning. When histogram binning is used in Tsallis GMI, as in Fig. \ref{rotation_surfaces}e, with 6 bits binning, we have no local maxima except for the solution, which benefits the gradient with smooth isosurfaces.

\subsection{Scale}
\begin{figure*}
	\centering
	\subcaptionbox{}{\includegraphics[scale=\myscale]{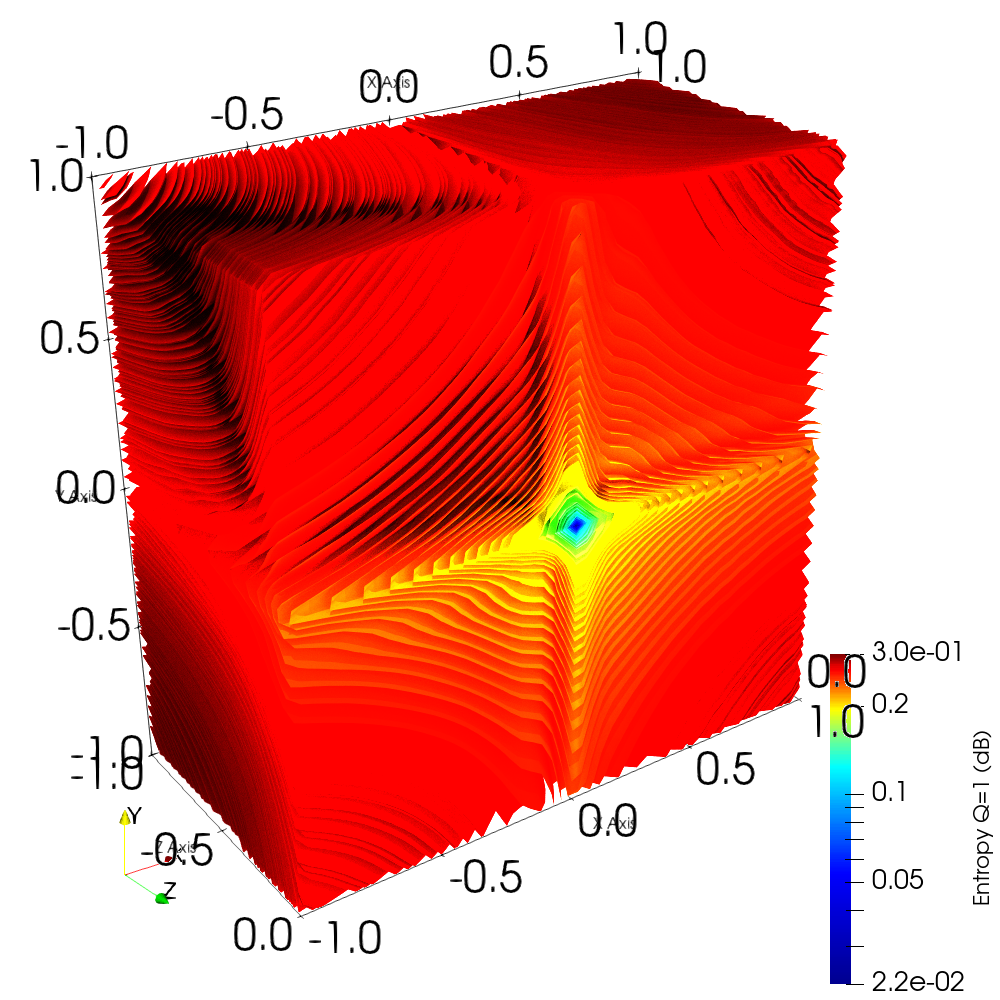}}%
	\hfill
	\subcaptionbox{}{\includegraphics[scale=\myscale]{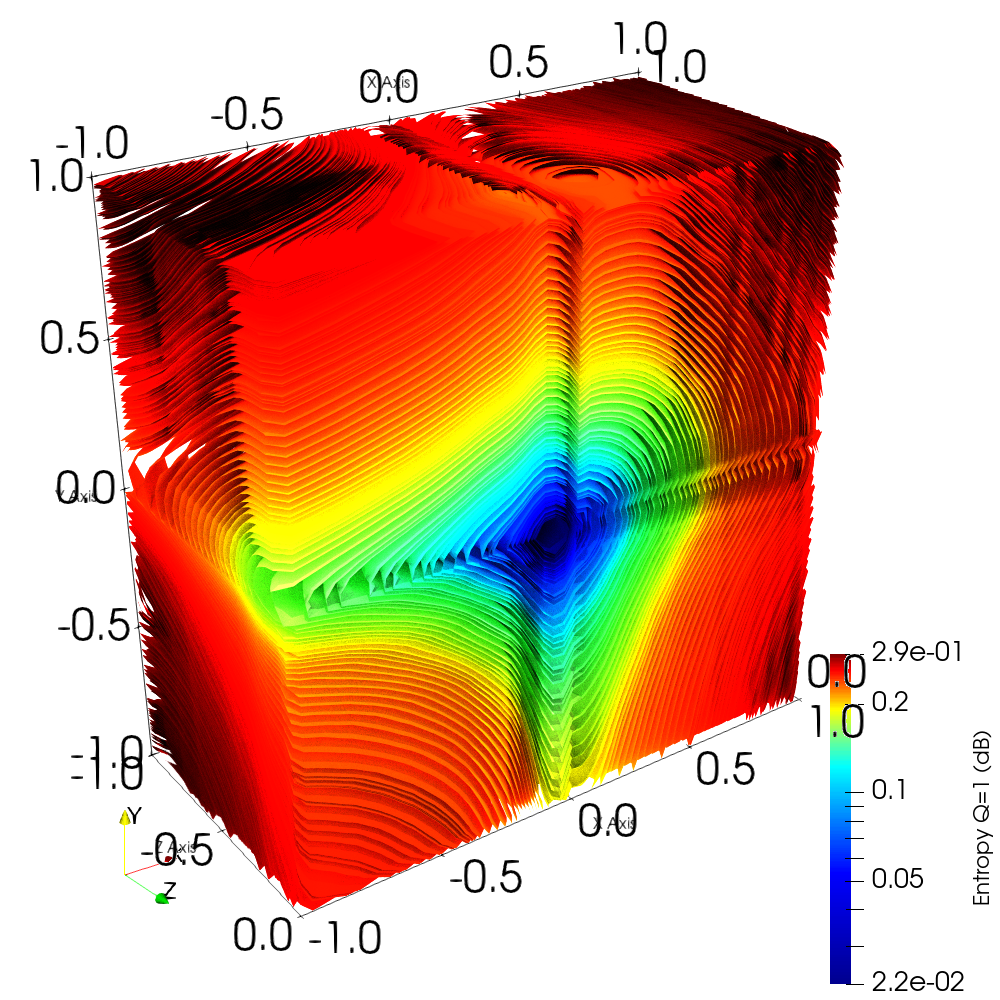}}%
	\hfill
	\subcaptionbox{}{\includegraphics[scale=\myscale]{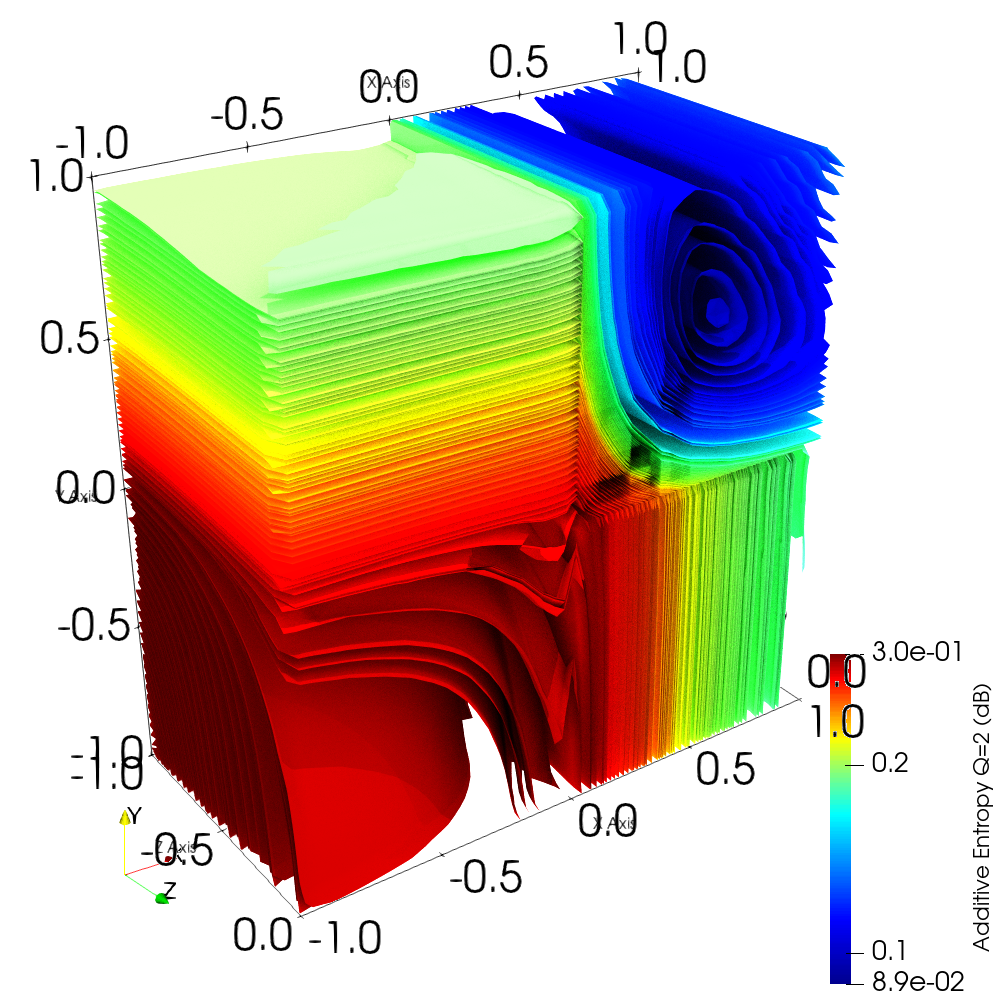}}%
	\hfill
	\subcaptionbox{}{\includegraphics[scale=\myscale]{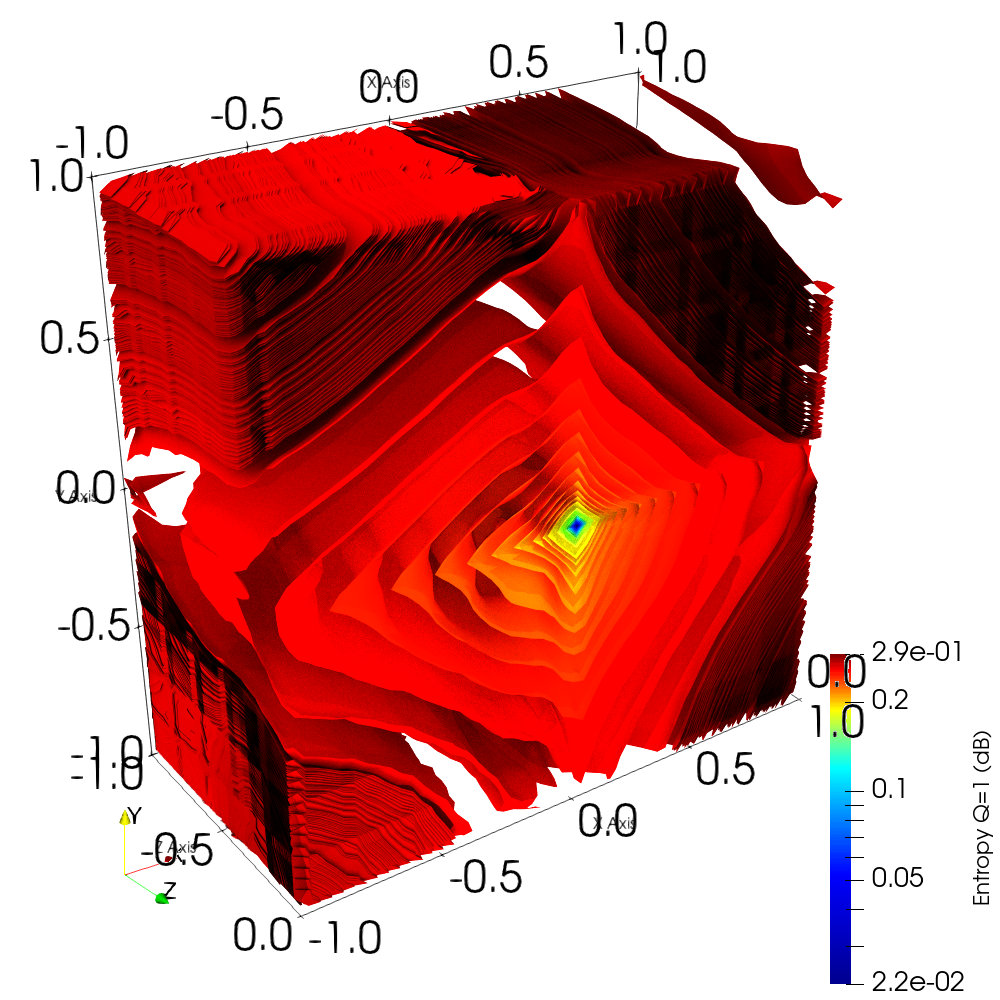}}%		
	\hfill
	\subcaptionbox{}{\includegraphics[scale=\myscale]{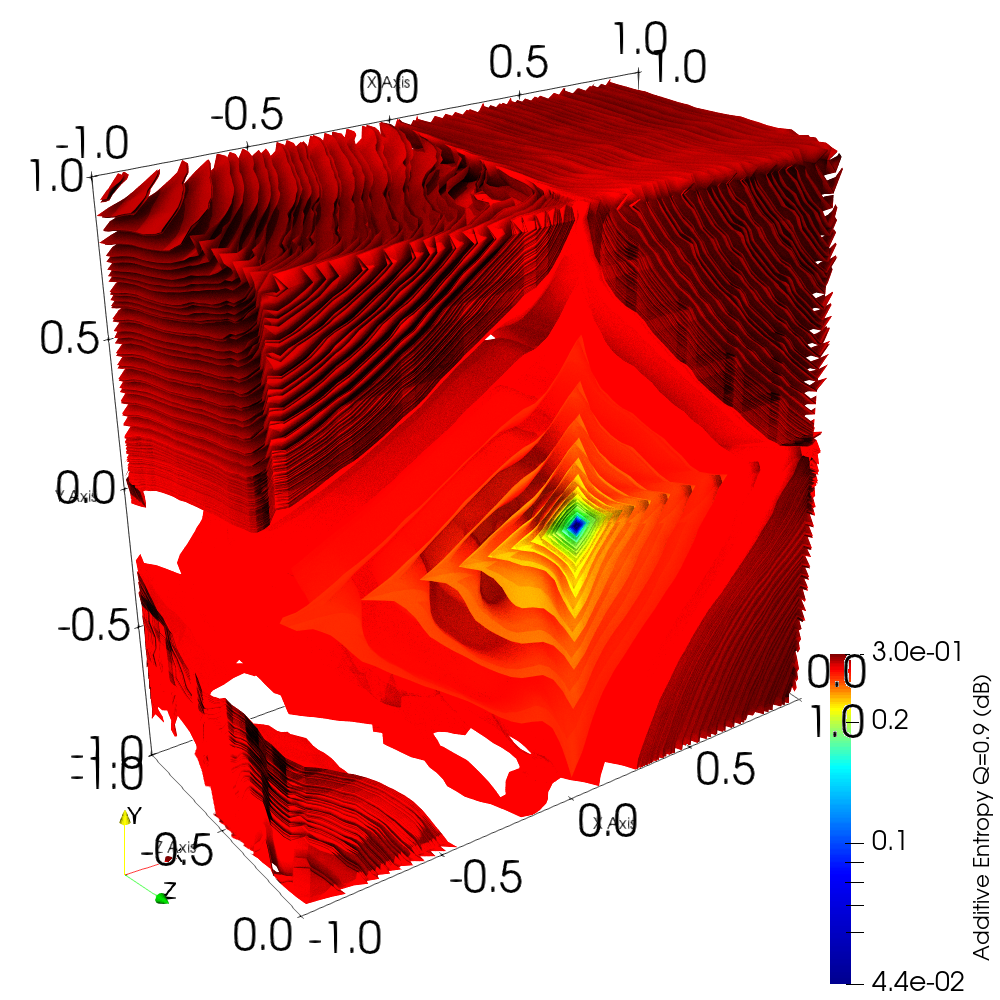}}%		
	\hfill
	\caption{\label{scale_surfaces}Isosurfaces of the MI function using scale parameters: (a) Mattes (b) Shannon (c) Tsallis GMI additive q=2.0 (d) Shannon with 10 bits binning (e) Tsallis GMI additive q=0.9 10 bit binning}
\end{figure*}	

Scale transform (Fig. \ref{scale_surfaces}) performs well using Mattes. Tsallis GMI, by itself, performs poorly (Fig. \ref{scale_surfaces}c), in additive or nonadditive form. However, using histogram bins (10 bits), as in the rotation transform case, improves Tsallis GMI and Shannon MI to compete with Mattes. Still, we need to use additive Tsallis GMI with the entropic index near Shannon, $q=0.9$ performs well, and $q=1.1$ works poorly in the situation. Therefore, using Tsallis GMI in additive or nonadditive form yields poor performance and will constrain the registration.  There is a possibility of using Tsallis GMI with an entropic index near Shannon, i.e., $q \to 1$, but it does not make sense since Shannon is a better alternative histogram binning. The Mattes outperforms Shannon, but both MI functions can register all the space and do not present local minima points other than the solution.

\subsection{Skew}
\begin{figure*}[!htpb]
	\centering
	\subcaptionbox{}{\includegraphics[scale=\myscale]{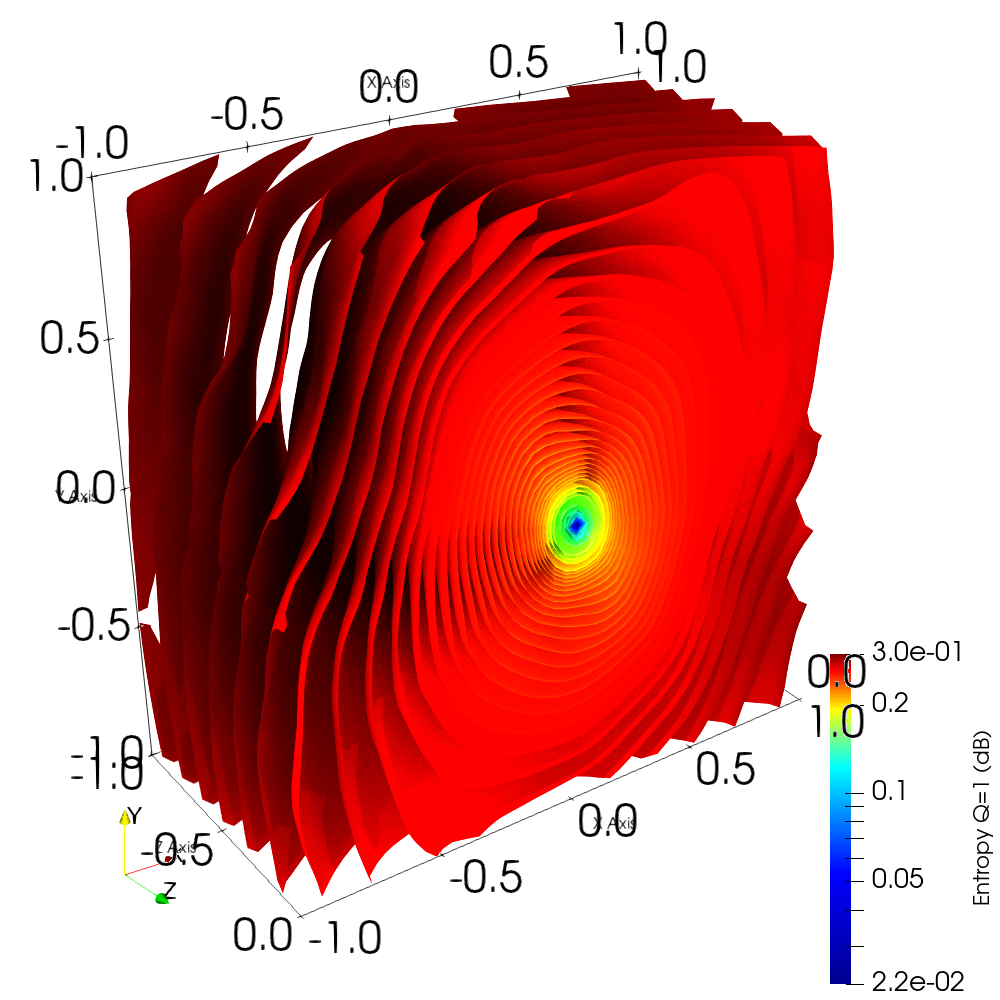}}%
	\hfill
	\subcaptionbox{}{\includegraphics[scale=\myscale]{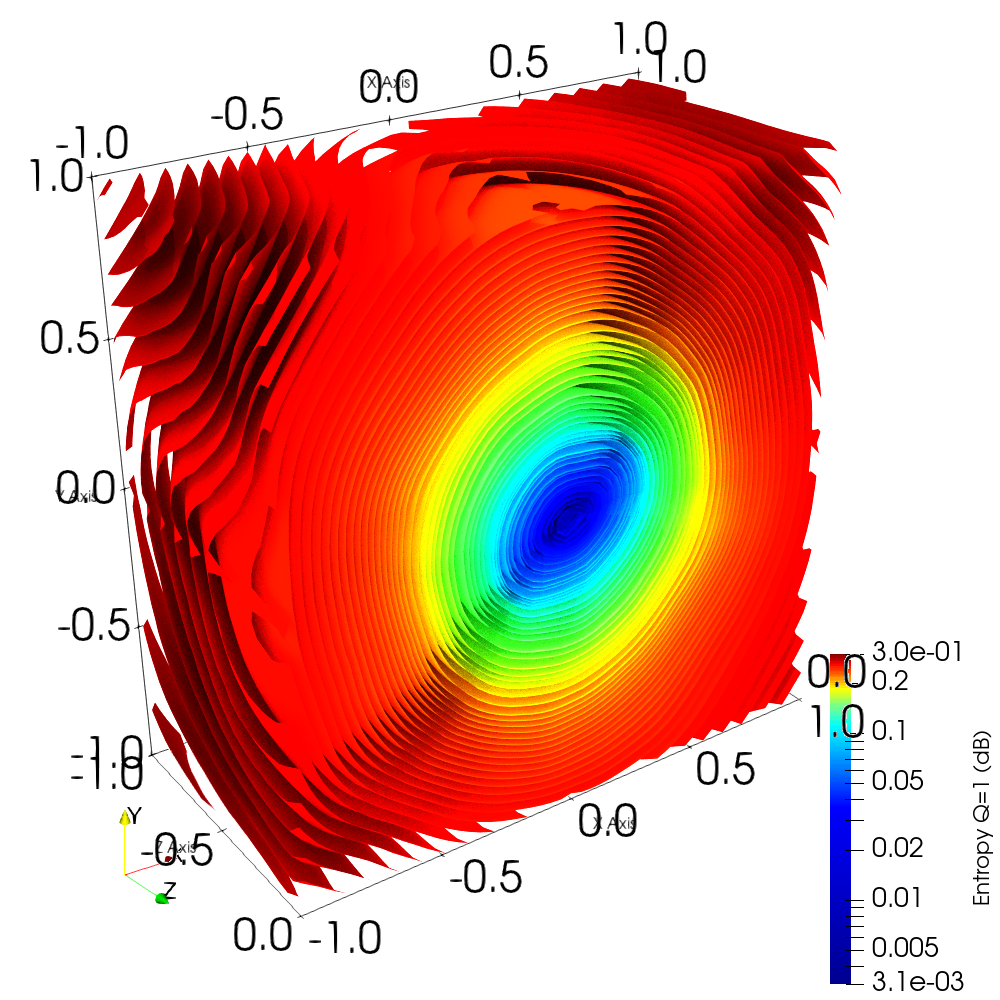}}%
	\hfill
	\subcaptionbox{}{\includegraphics[scale=\myscale]{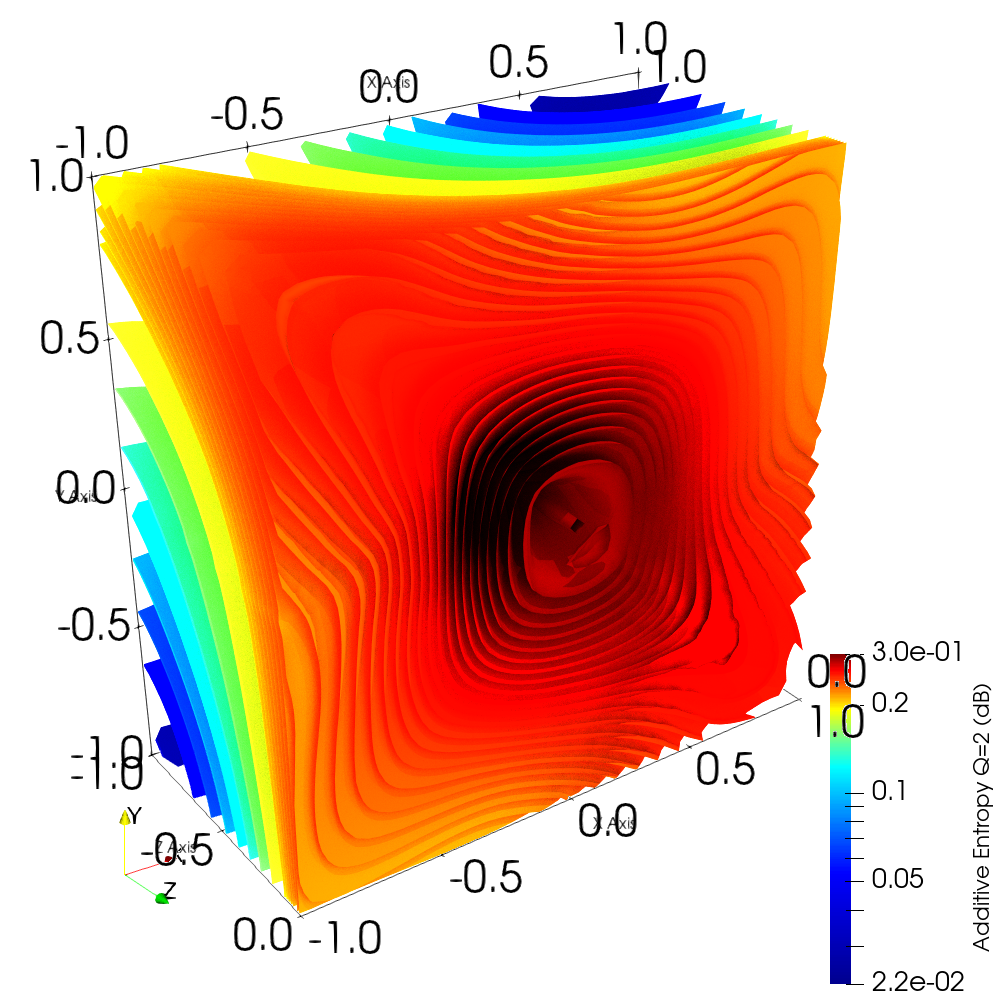}}%
	\hfill
	\subcaptionbox{}{\includegraphics[scale=\myscale]{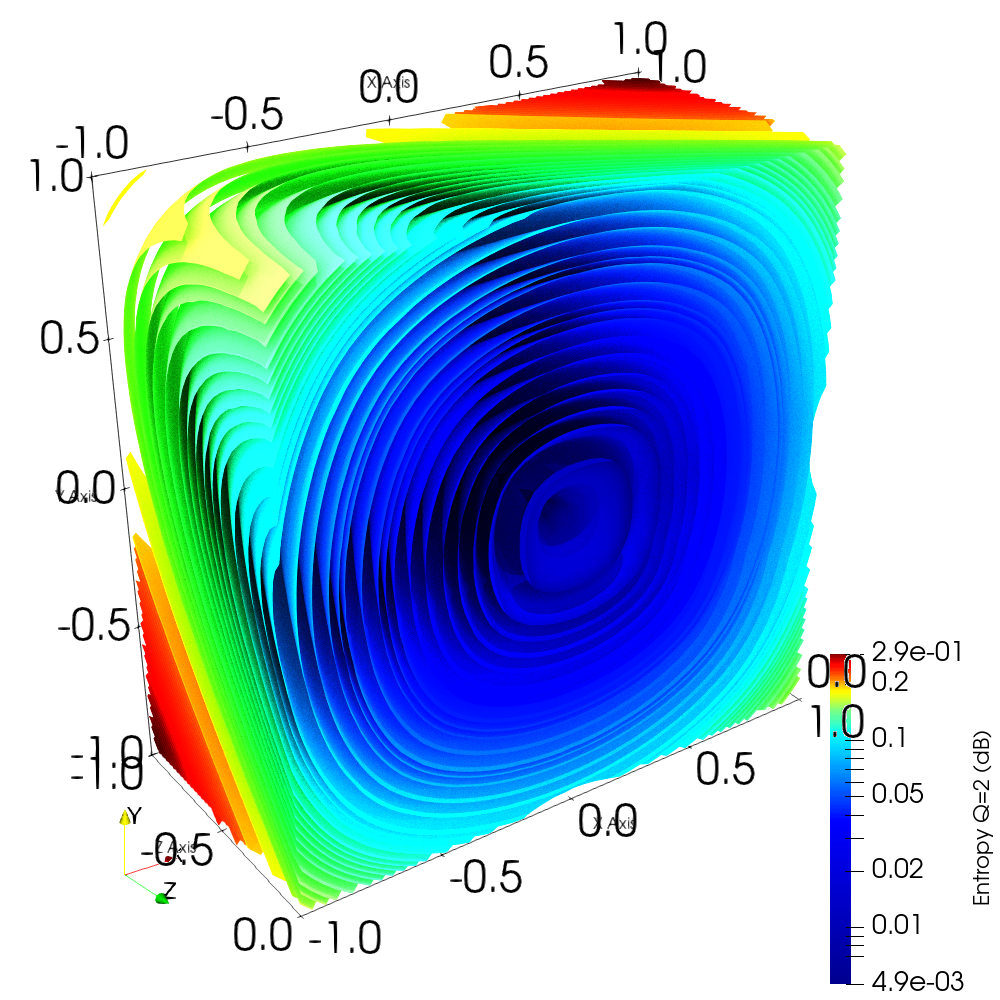}}%
	\subcaptionbox{}{\includegraphics[scale=\myscale]{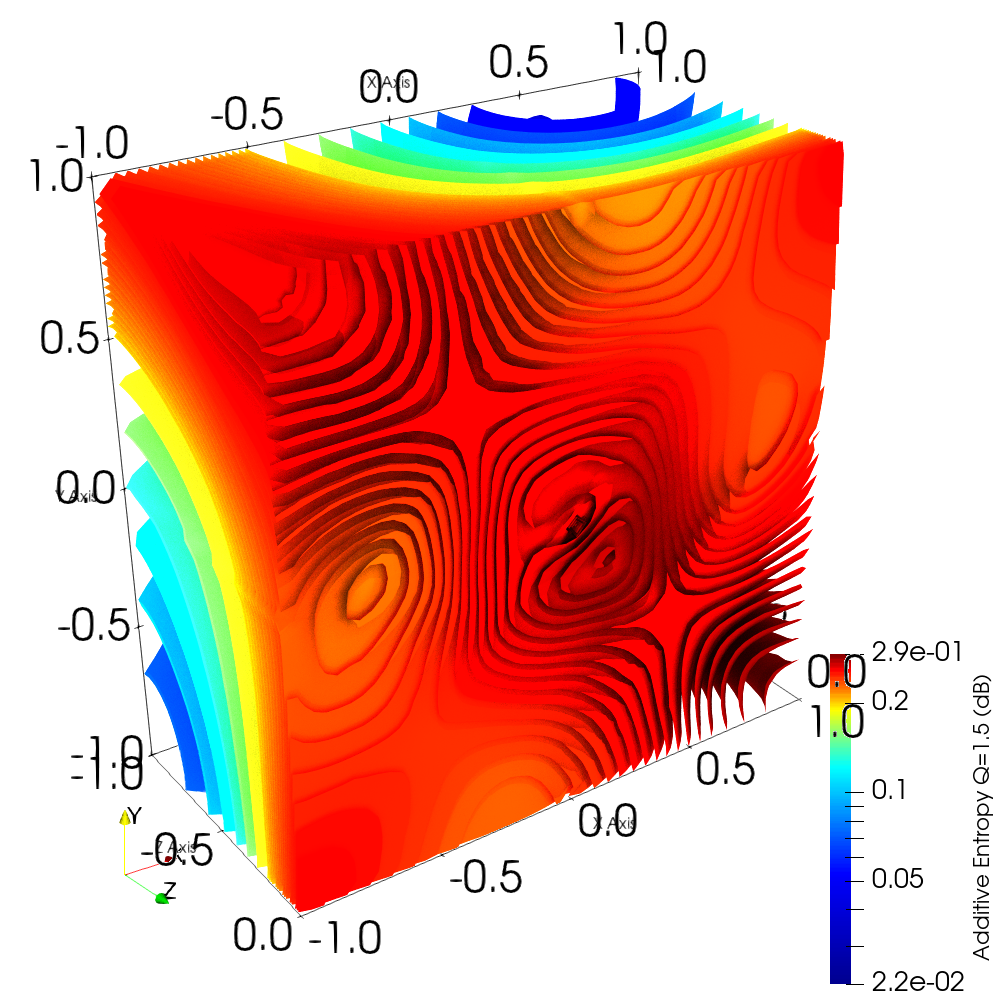}}%
	\hfill
	\caption{\label{skew_surfaces}Isosurfaces of the MI function using skew parameters: (a) Mattes, (b) Shannon, (c) Tsallis GMI additive q=2.0, (d) Tsallis GIM nonadditive q=2.0, and (e) Tsallis GMI additive q=1.5}
\end{figure*}	

Skew transforms isosurfaces shown in Fig. \ref{skew_surfaces} indicate good performance for both Mattes and Shannon. However, for Tsallis, one must use the nonadditive form for better performance. In the additive Tsallis GMI, only using $q \geq 2$ performed well, while for nonadditive Tsallis GMI, all $q > 1$ performed well. 

\subsection{Registration Simulation}
\begin{table*}[t]
	\centering
	\begin{tabular}{lccrrrr}
		\toprule
    Metric & Q & Binning & Translation & Rotation & Scale & Skew \\
		\midrule
    Mattes & 1.0 & Yes & 65.81\% & 90.94\% & \textbf{99.99\%} & \textbf{99.99\%} \\
    Shannon & 1.0 & No & 98.08\% & 87.26\% & \textbf{99.99}\% & \textbf{99.99}\% \\
    Tsallis & 0.5 & No & 0.2\% & 18.46\% & 41.41\% & 18.86\% \\
    Tsallis & 1.1 & 12 bit & \textbf{99.99\%} & 72.84\% & 64.87\% & 99.99\% \\
    Tsallis & 1.5 & No & \textbf{99.99\%} & 62.06\% & 68.14\% & 99.96\% \\
    Tsallis Additive & 0.5 & No & 8.29\% & 65.86\% & 77.93\% & 96.51\% \\
    Tsallis Additive & 1.5 & No & 86.14\% & 7.07\% & 47.38\% & 52.26\% \\
    Tsallis Additive & 0.7 & 12 bit & 29.96\% & \textbf{97.58\%} & 84.16\% & 99.98\% \\
    Tsallis Additive & 1.8 & 12 bit & 84.47\% & 45.29\% & \textbf{99.99\%} & \textbf{99.99\%} \\
		\bottomrule
	\end{tabular}
	\caption{Comparison of different metrics and the percentage of the studied space that each metric can register; Q is the entropic index, best results in bold}
	\label{table:full_metrics}
\end{table*}	

The successful registration rates of each metric and transform are presented in Table \ref{table:full_metrics}. The rates are estimated starting at the center voxel of images generated in section \ref{sec:experimental_visualization}, and growing by checking whether neighbors' elements are higher or lower, to minimize or maximize the metric, simulating a gradient descent (or ascent) algorithm. This will not guarantee the same results in a real optimizer since gradient descent's learning rate can take different paths than the one we are simulating. However, we can have some estimative of the registration quantitative capacity of each metric and transform pair. Tsallis GMI outperforms other translation metrics without any histogram binning. Mattes has good performances in the other transforms. One can see from Fig. \ref{translation_surfaces} and Table \ref{table:full_metrics} that Mattes has problems in translation. Shannon MI has a fair result, with only its rotation performing marginally worse than Mattes.

Tsallis GMI functions show superiority on rotation when using histogram binning, as shown in the results using 12 bits binning. However, one must be careful in mixing the entropic index and additive and nonadditive forms in different transforms. Each parameter space should have a different $q$ entropic index and different additivity features to produce successful registration. These parameters space features can explain why Tsallis GMI entropy image registration results reported in previous literature are controversial and not consistent. The typical approach uses a unique $q$ entropic index and technique over the entire parameter space. This approach can show spurious results depending on the used methodology, e.g., one Tsallis GMI settings may be useful for the translation but fail to register rotations in the same scenario.

\subsection{Monte Carlo}

\begin{figure*}
  \centering\includegraphics[width=\textwidth]{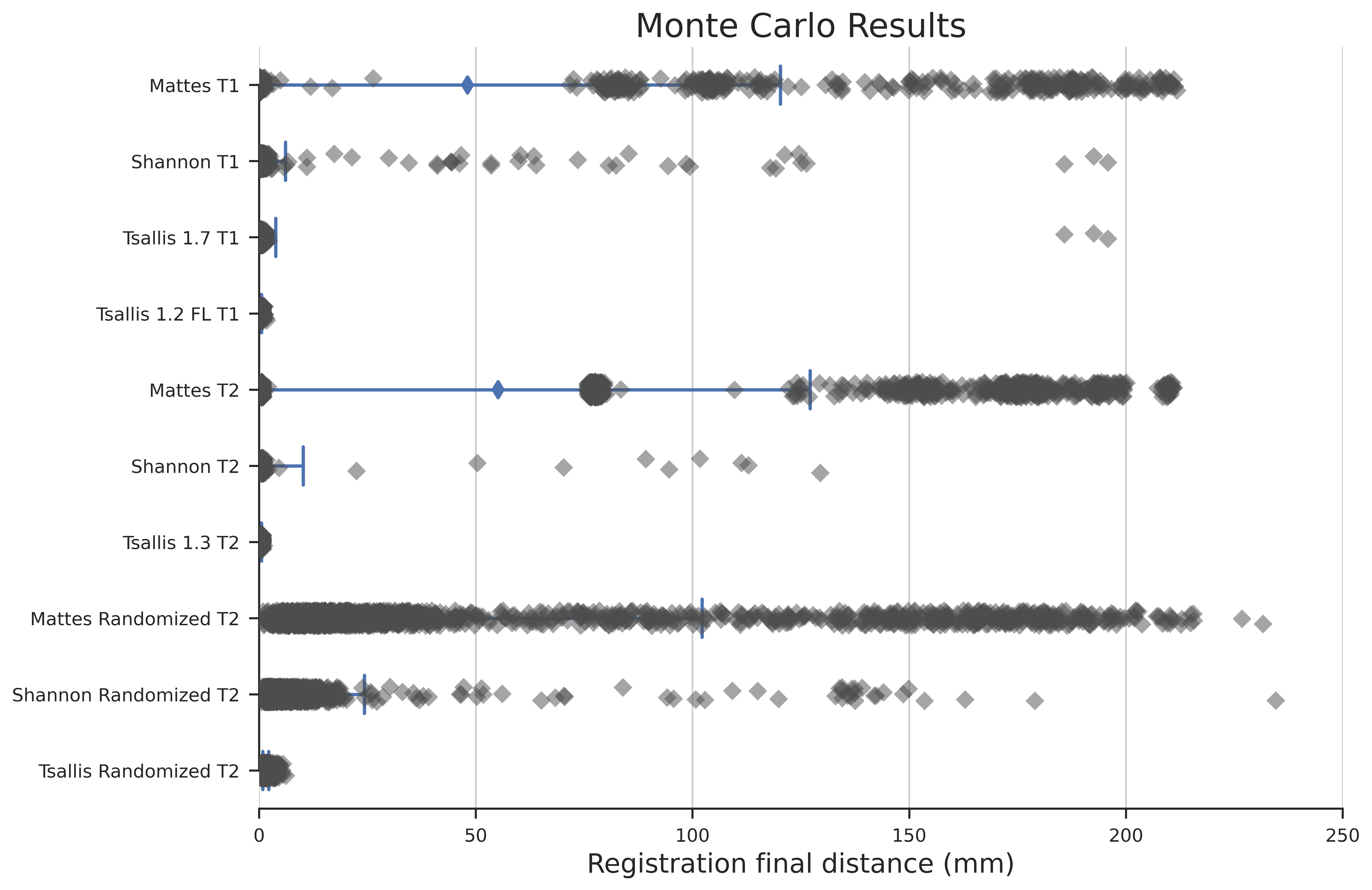}
  \caption{\label{fig:monte_carlo_results}Monte Carlo registration essays for multiple methods and scenarios, each gray diamond represents a single registration result, the blue line is the standard deviation range (68\%) and the blue diamond is the mean distance of the final registration parameters (close to the left is better).}
\end{figure*}

Figure \ref{fig:monte_carlo_results} shows a scatter plot for the Monte Carlo registration experiments. The scatter plots present the final distances for randomized initial distances. Mattes has a poor performance in all scenarios, with the image registration's end parameters far from the center and some regions clearly showing local minima problems. An example is the Mattes T2 scenarios, where we have a local minimum about the 75 mm distance,  also shown in the Mattes T1 scenario. These problems reflect on the results' statistical properties, with a mean far from the center and a large deviation.

One can see a much better result in all scenarios for the proposed methods, i.e., Tsallis GMI outperforms Mattes and even Shannon. Shannon  MI  produces more failed results than Tsallis GMI in Monte Carlo experiments. The failure rates are even more explicit in the Randomized T2 experiments. It is worth noting that the Randomized T2 is the most critical and challenging scenario for the similarity metrics, forcing them to register a T1 image with a T2 image from random subjects, meaning that in most cases, the T1 and T2 subjects are different.

Another result from Figure \ref{fig:monte_carlo_results} is the usage of a heavier interpolation in terms of computational needs. The Tsallis GMI 1.2 FL uses a Lanczos algorithm with fast mathematics for trigonometric functions, i.e., the FastLanczos algorithm. We have slightly better results in comparison to the normal one that uses nearest-neighbor interpolation.

Table \ref{table:monte_carlo_results} presents results showing the success rate of the similarity metrics depending on the acceptable final distance, i.e., the thresholds for a successful or failed registration. Mattes has a poor performance in the Randomized T2 scenario, even with an acceptable distance of $5 mm$. We obtained $5.05\%$ success for Mattes, $56.50\%$ for Shannon, and $99.75\%$ for Tsallis GMI using $q$ entropic index of $1.3$ with the $5 mm$ distance threshold.

\begin{figure*}
  \begin{center}
  \includegraphics[width=0.3\textwidth]{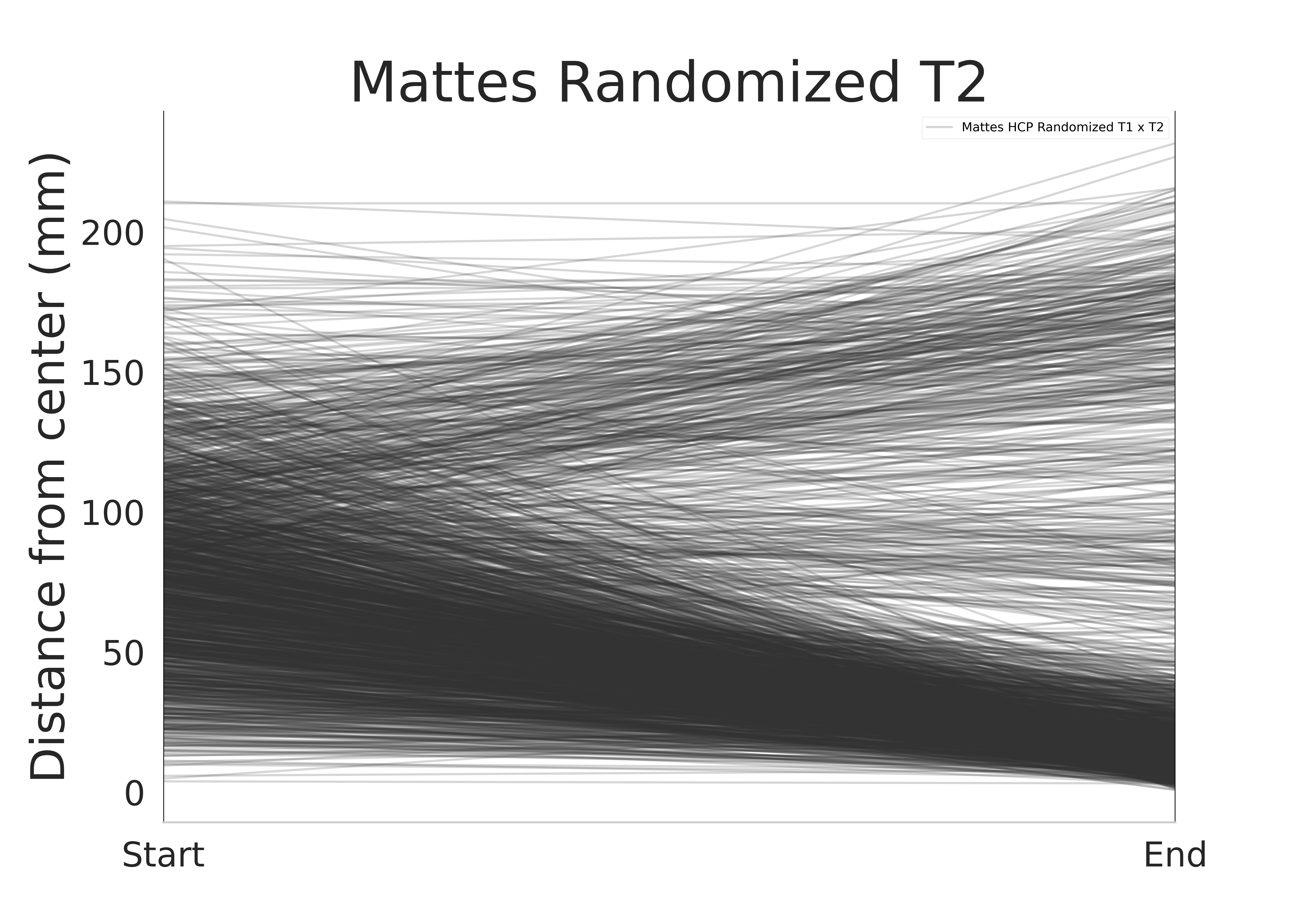}
  \includegraphics[width=0.3\textwidth]{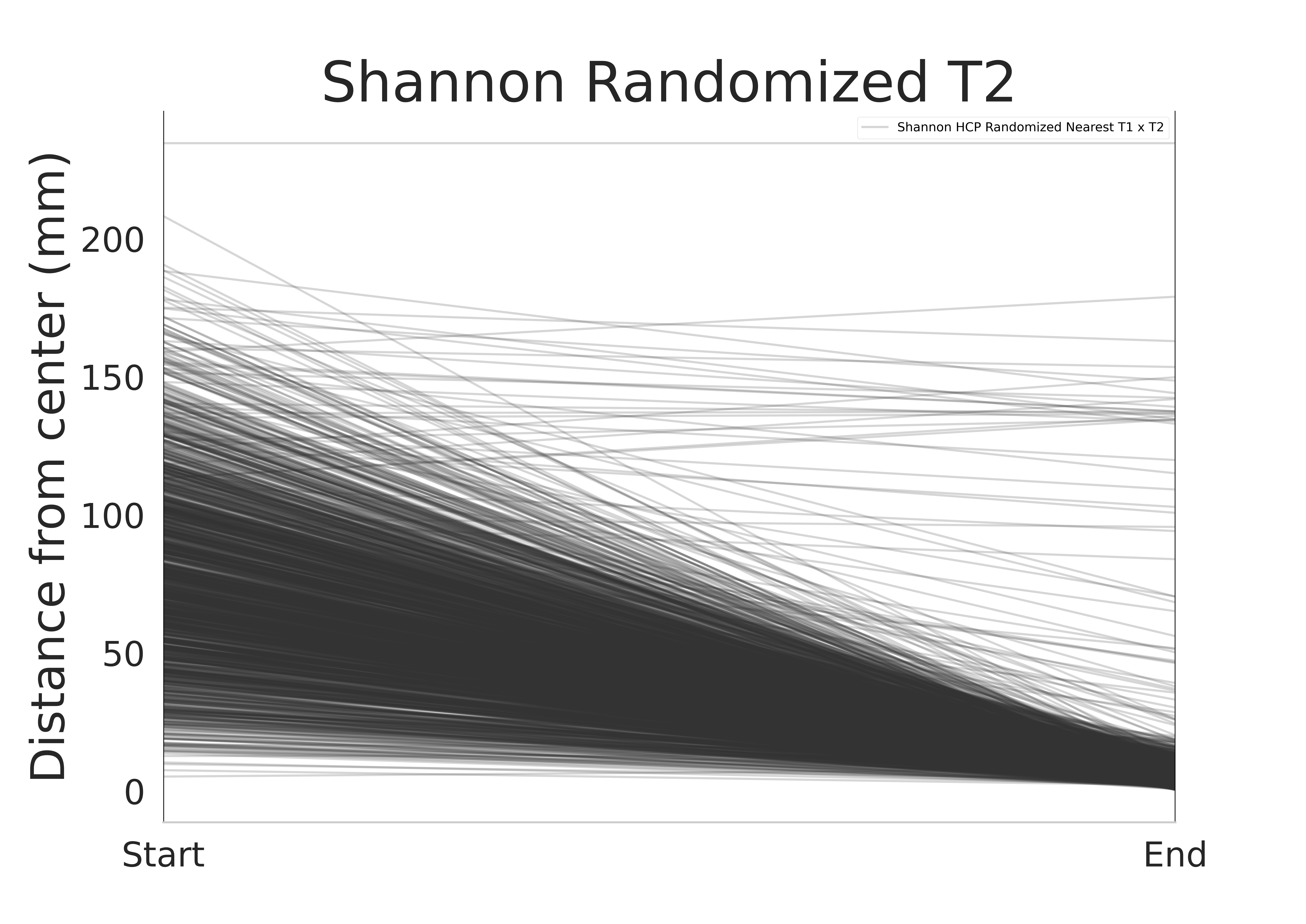}
  \includegraphics[width=0.3\textwidth]{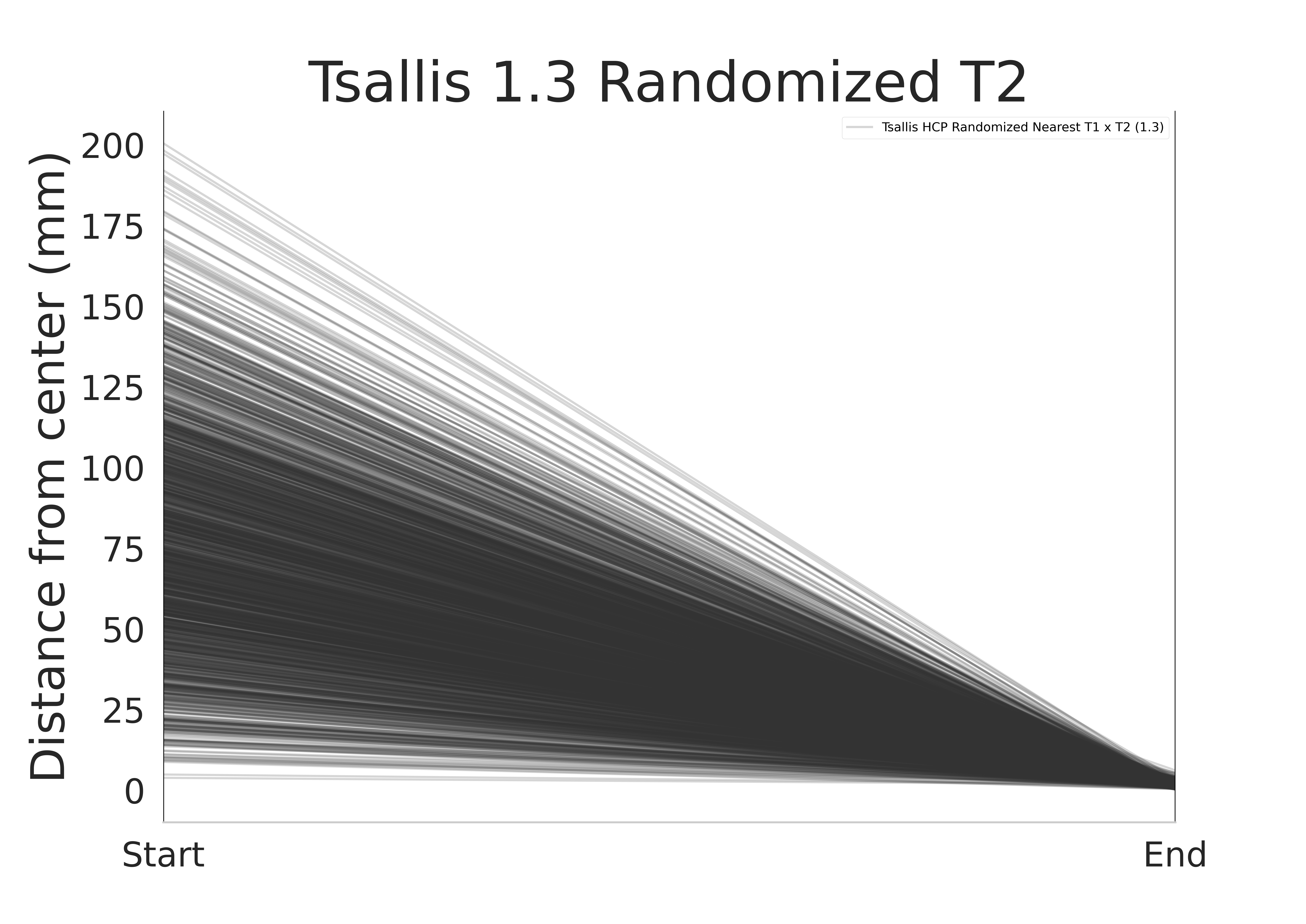}
\end{center}
\caption{\label{fig:monte_carlo_parallel}Monte Carlo registration essays as parallel coordinates, each line represent a registration essay where start is the distance from the start parameters set and end is the final distance from the optimal registration, an ideal registration will come from anywhere on the left and end on the right bottom}
\end{figure*}

Figure \ref{fig:monte_carlo_parallel} shows results from the experiments using a parallel coordinates plot with the starting distances of each random sample and the corresponding end distance, i.e., the final registration distances. The Figure indicates the superiority of Tsallis in this scenario since all starting distances converge to the center position at the end, i.e., small distances. In contrast, we have some registrations leading to worse distances than it started for Shannon, and on a different level, we have many for Mattes.

\begin{table*}
  \centering\begin{tabular}{llllrrr}
    \toprule
 & & & & \multicolumn{3}{c}{Acceptable distance (\%)} \\
 \cmidrule{5-7}%
    Method & Scenario & Mean & Deviation & 1mm  & 3mm & 5mm \\
\midrule

    Mattes & T1 & 48.08 & 72.24 & 64.70 & 65.80 & 66.00 \\
    Shannon Nearest & T1 & 0.68 & 5.38 & 97.04 & 99.61 & 99.62 \\
    Tsallis Nearest (1.7) & T1 & 0.48 & 3.31 & 99.33 & 99.97 & 99.97 \\
    Tsallis FastLanczos (1.2) & T1 & 0.39 & 0.15 & 98.90 & 100.00 & 100.00 \\
    \midrule
    Mattes & T2 & 55.14 & 72.01 & 58.15 & 58.25 & 58.25 \\
    Shannon Nearest & T2 & 1.48 & 8.69 & 97.20 & 99.00 & 99.10 \\
    Tsallis Nearest (1.3) & T2 & 0.45 & 0.10 & 99.90 & 100.00 & 100.00 \\
    \midrule
    Shannon Nearest & Randomized T1 & 1.62 & 10.88 & 94.09 & 98.21 & 98.26 \\
    Tsallis Nearest (1.3) & Randomized T1 & 0.63 & 6.04 & 98.87 & 99.91 & 99.91 \\
     \midrule
    Mattes & Randomized T2 & 45.25 & 56.94 & 0.10 & 1.10 & 5.05 \\
    Shannon Nearest & Randomized T2 & 7.46 & 16.81 & 2.10 & 37.15 & 56.50 \\
    Tsallis Nearest (1.3) & Randomized T2 & 1.51 & 0.71 & 20.85 & 95.80 & 99.75 \\

\bottomrule
  \end{tabular}
  \caption{\label{table:monte_carlo_results}Monte Carlo results, mean and deviation in mm, acceptable distance is the percentage of registrations within this distance from the center}
\end{table*}

\section{Discussion}
This study assessed Tsallis GMI as a similarity metric working as a cost function optimized in medical image registration compared to Shannon and Mattes. We presented the cost function isosurfaces evaluations for translation, rotation, scaling, and skewness with favorable GMI function outcomes.

The studies discussed in section \ref{sec:intro_history} had been compromised the  MI cost function due to a lack of computational power. To compute all voxels with high precision would take too much time. 
The emerge of parallel computing allowed, with high bandwidth GPUs, to sweep all the translation space as in section \ref{sec:experimental_visualization} within one hour, making the experiments included in this study possible.

In this paper, we exploited the current computational power to assess and understand how MI is affected by the geometric transforms and different images, in our case MRI T1 and T2. The study's importance is that a metric with a better mathematical image yields better results when optimized by helping the optimizer not fall in local extrema points, and having a quasiconvex or quasiconcave allows a gradient descent optimizer to find the solution. A quasiconvex function has the property that the local minimum is the global minimum, so in almost all cases, the gradient descent will reach our global minimum point and successfully register the image.

Although we achieved good results using Tsallis MI functions, the entropic index and additivity demand a sensible tunning and further study depending on the images used. From Fig. \ref{translation_surfaces}, we can see the difference where Tsallis Additive has several local minima, and the nonadditive function has a single local minimum that is a global minimum, i.e., the function is quasiconvex. Fig. \ref{rotation_surfaces} shows a different scenario with both Tsallis functions having a local minimum, forcing us to apply histogram binning to have the quasiconvex space. Fig. \ref{skew_surfaces} exemplify the impact of the entropic index where subfigures (d) and (e) have, respectively, $q=2.0$ and $q=1.5$, and (d) shows a quasiconvex space where (e) shows local minimum points. 

The results from Monte Carlo were auspicious (Table \ref{table:monte_carlo_results}), reaching an ideal score if we consider results within $3mm$ of distance acceptable on the scenario of T1 and T2 from the same patients. One can see a real problem with the current ITK Mattes implementation, i.e.,  the hardest scenario where we randomized the patients, having $5.05\%$ success within $5mm$ where Shannon achieves $56.5\%$ and Tsallis GMI $99.75\%$ success rate. The difference in registration quality is more detailed in Fig. \ref{fig:monte_carlo_results}, where we can see the scatter plot showing all the Monte Carlo trials results. Mattes have several registration results above the $50mm$ region, and one can observe some local minimum results for Mattes T2 around $75mm$. The Monte Carlo results for Tsallis GMI are close to $0mm$. Tsallis GMI results usually have no outliners, i.e., failed registration, except for the few essays with Tsallis GMI 1.7 T1. Fig. \ref{fig:monte_carlo_parallel} compares the Randomized T2 results where one can better see the problem with Mattes registration results farthest from the starting points. One can also observe Shannon having a few problems and Tsallis GMI with an outstanding performance. 

\section{Conclusion}
In this paper, we analyzed GMI, MI, and Mattes similarity metric functions using 3D images and isosurfaces contours, allowing a better view of the local extrema points that challenge medical image registration. 
Besides, we developed a technique for image registration using Tsallis Mutual Information, using various additive and nonadditive forms, and multiple entropic indexes over the parameter space.

Moreover, the analysis of Tsallis nonadditive GMI equation \eqref{eq:MI_normal}, ignoring the formality of mutual entropy in physics, with its outputs varying with transform parameters, shows that using the classic Shannon MI equation with Tsallis entropy, called in this paper nonadditive Tsallis, may benefit in some scenarios. 
In contrast, in others, we need to use the Tsallis additivity. 

Our results suggest using a specific $q$ value for each of the parameters elements and also alternating between additive and nonadditive Tsallis depending on the parameter being optimized. To better explain, each of the transforms groups, translation, rotation, scale and skew, needs its own value of $q$ and setting of additive or nonadditive Tsallis GMI.
From our experimental data, for translation you want $q>1$ and nonadditive, for rotation $q>1$ and additive, for scale Shannon ($q=1$) is a better alternative and for skew $q>1$ and nonadditive.
Although we only assessed translation through Monte Carlo simulations, one can reason toward similar features from the remaining metrics, i.e., rotation and skewness. Furthermore, scaling results argue weakly toward GMI, with effectiveness very close to the investigated classic metrics.  

Our experiments' major result is that we need to rethink the computational approach where we do not need to take shortcuts like voxel sampling or force a histogram binning. The current computational power allows exhaustive scanning of the available data, having a better overall registration assessment. Histogram binning still has its place, as in the rotation transforms, necessary to make the space quasiconvex.

We hope further studies can benefit from the analysis tools used in this paper, like the 3D isosurfaces, and inspect the metric's space to evolve it, helping the optimizer achieve a better registration. 
Although, in some techniques, like deformable registration, we have a higher number of parameters that challenges the visualization, having simulations of how the metric behave concerning the parameters can help the researcher improve the metric or even choose a more suitable optimizer for the metric space.

\section*{CRediT author statement}
\textbf{Vinicius Pavanelli Vianna:} Conceptualization, Methodology, Software, Validation, Formal analysis, Investigation, Resources, Data Curation, Writing - Original Draft, Visualization. 
\textbf{Luiz Otavio Murta Jr.:} Conceptualization, Methodology, Resources, Writing - Reviewer \& Editing, Supervision, Project administration, Funding acquisition.

\section*{Acknowledgments}
        Data were provided by the Human Connectome Project, WU-Minn Consortium (Principal Investigators: David Van Essen and Kamil Ugurbil; 1U54MH091657), funded by the 16 NIH Institutes and Centers that support the NIH Blueprint for Neuroscience Research; and by the McDonnell Center for Systems Neuroscience at Washington University.

%%Harvard
\bibliographystyle{model2-names.bst}\biboptions{authoryear}
\bibliography{jabref}

\end{document}